\documentclass[aps,prl,twocolumn,superscriptaddress,longbibliography]{revtex4-1}
\pdfoutput=1

\usepackage{graphicx} 
\usepackage{amssymb}
\usepackage{amsmath}
\usepackage{mathtools}
\usepackage{color}
\usepackage{ifthen}
\usepackage{braket}

\usepackage{color}
\usepackage[colorlinks,bookmarks=false,citecolor=darkblue,linkcolor=red,urlcolor=blue]{hyperref}

\definecolor{darkred}{rgb}{0.7,0.0,0.0}
\definecolor{darkblue}{rgb}{0,0.02,0.45}

\usepackage[capitalise]{cleveref}

\newcommand{\GammaMag}[1][]{\ensuremath{\Gamma_{\mathrm{B#1}}}}

\newcommand{\aRuCl}{$\alpha$-RuCl$_3$}
\newcommand{\Gruneisen}{Gr\"uneisen}
\newcommand{\Ctot}{\ensuremath{C_{\mathrm{tot}}}}
\newcommand{\Csample}{\ensuremath{C_{\mathrm{Sa}}}}
\newcommand{\Cplatform}{\ensuremath{C_{\mathrm{Cell}}}}
\newcommand{\dMdT}[1][]{\ensuremath{\ensuremath{\partial{}M_{\mathrm{#1}}/\partial{}T}}}
\newcommand{\dSdB}[1][]{\ensuremath{\ensuremath{\partial{}S_{\mathrm{#1}}/\partial{}B}}} 
\newcommand{\Bc}[1][]{\ensuremath{B_{\mathrm{c#1}}}}

\newcommand{\control}{\ensuremath{\lambda}}

\renewcommand{\Bc}[1][]{\ifthenelse{\equal{#1}{3}}{\ensuremath{B^\ast}}{\ensuremath{B_c^{\mathrm{AF#1}}}}}
\newcommand{\GammaH}[1][]{\ensuremath{\Gamma_{\mathrm{B#1}}}}

\newcommand{\rucl}{$\alpha$-RuCl$_3$}

\begin{document}

\title{Thermodynamic perspective on field-induced behavior of $\alpha$-RuCl$_3$}

\author{S. Bachus}
\email[]{sebastian.bachus@physik.uni-augsburg.de}
\affiliation{Experimental Physics VI, Center for Electronic Correlations and Magnetism, University of Augsburg, Universitätsstr.~1, 86159 Augsburg, Germany}
\author{D.~A.~S.~Kaib}
\email[]{kaib@itp.uni-frankfurt.de}
\affiliation{Institute of Theoretical Physics, Goethe University Frankfurt, Max-von-Laue-Strasse 1, 60438 Frankfurt am Main, Germany}

\author{Y. Tokiwa}
\affiliation{Experimental Physics VI, Center for Electronic Correlations and Magnetism, University of Augsburg, Universitätsstr.~1, 86159 Augsburg, Germany}

\author{A. Jesche}
\affiliation{Experimental Physics VI, Center for Electronic Correlations and Magnetism, University of Augsburg, Universitätsstr.~1, 86159 Augsburg, Germany}

\author{V. Tsurkan}
\affiliation{Experimental Physics V, Center for Electronic Correlations and Magnetism, University of Augsburg, Universitätsstr.~1, 86159 Augsburg, Germany}
\affiliation{Institute of Applied Physics, Chisinau MD-2028, Republic of Moldova}

\author{A.~Loidl}
\affiliation{Experimental Physics V, Center for Electronic Correlations and Magnetism, University of Augsburg, Universitätsstr.~1, 86159 Augsburg, Germany}

\author{S.~M.~Winter}
\email[]{winter@physik.uni-frankfurt.de}
\affiliation{Institute of Theoretical Physics, Goethe University Frankfurt, Max-von-Laue-Strasse 1, 60438 Frankfurt am Main, Germany}

\author{A. A. Tsirlin}
\email[]{altsirlin@gmail.com}
\affiliation{Experimental Physics VI, Center for Electronic Correlations and Magnetism, University of Augsburg, Universitätsstr.~1, 86159 Augsburg, Germany}

\author{R.~Valent\'\i}
\email[]{valenti@itp.uni-frankfurt.de}
\affiliation{Institute of Theoretical Physics,
Goethe University Frankfurt,
Max-von-Laue-Strasse 1, 60438 Frankfurt am Main, Germany}

\author{P. Gegenwart}
\email[]{philipp.gegenwart@physik.uni-augsburg.de}
\affiliation{Experimental Physics VI, Center for Electronic Correlations and Magnetism, University of Augsburg, Universitätsstr.~1, 86159 Augsburg, Germany}


\begin{abstract}
Measurements of the magnetic Gr\"uneisen parameter ($\Gamma_{\rm B}$) and specific heat on the Kitaev material candidate $\alpha$-RuCl$_3$ are used to access in-plane field and temperature dependence of the entropy up to 12\,T and down to 1\,K. No signatures corresponding to phase transitions are detected beyond the boundary of the magnetically ordered region, but only a shoulder-like anomaly in $\Gamma_{\rm B}$, involving an entropy increment as small as $10^{-5} R\log 2$.	These observations put into question the presence of a phase transition between the purported quantum spin liquid and the field-polarized state of $\alpha$-RuCl$_3$. We show theoretically that at low temperatures $\Gamma_{\rm B}$ is sensitive to crossings in the lowest excitations within gapped phases, and identify the measured shoulder-like anomaly as being of such origin. Exact diagonalization calculations 
demonstrate that the shoulder-like anomaly can be reproduced in extended Kitaev models that gain proximity to an additional phase at finite field without entering it. 
	We discuss manifestations of this proximity in other measurements.
\end{abstract}

\maketitle


Quantum spin liquids (QSLs) describe novel states of matter that violate
Landau's concept of broken symmetry and associated order parameters~\cite{wen}.
These states feature unconventional quasiparticle excitations, such as spinons,
Majorana fermions, or artificial photons~\cite{Balents2010}. For example, the
exactly solvable Kitaev model leads to a $Z_2$ QSL ground state with emergent
fractionalized Majorana excitations~\cite{Kitaev}. Recent efforts
focused on
compounds with heavy transition-metal elements as experimental realizations of
this model~\cite{Jackeli,rau2016spin,Winter2017,Takagi2019},
and several promising materials
including the two-dimensional
$\alpha$-RuCl$_3$, Na$_2$IrO$_3$ and $\alpha$-Li$_2$IrO$_3$
have been identified. Although interactions beyond the
Kitaev model cause long-range magnetic order in the above mentioned
materials, the presence of
a strong Kitaev exchange has been suggested to
 lead to proximate QSL behavior in the paramagnetic
state above the N\'eel temperature~\cite{Banerjee,Revelli2019,kim2020} and in applied magnetic
fields upon the suppression of the ordered phase~\cite{Janssen2019,lee2020magnetic}. 

\begin{figure}[t!]
\includegraphics[width=0.45\textwidth]{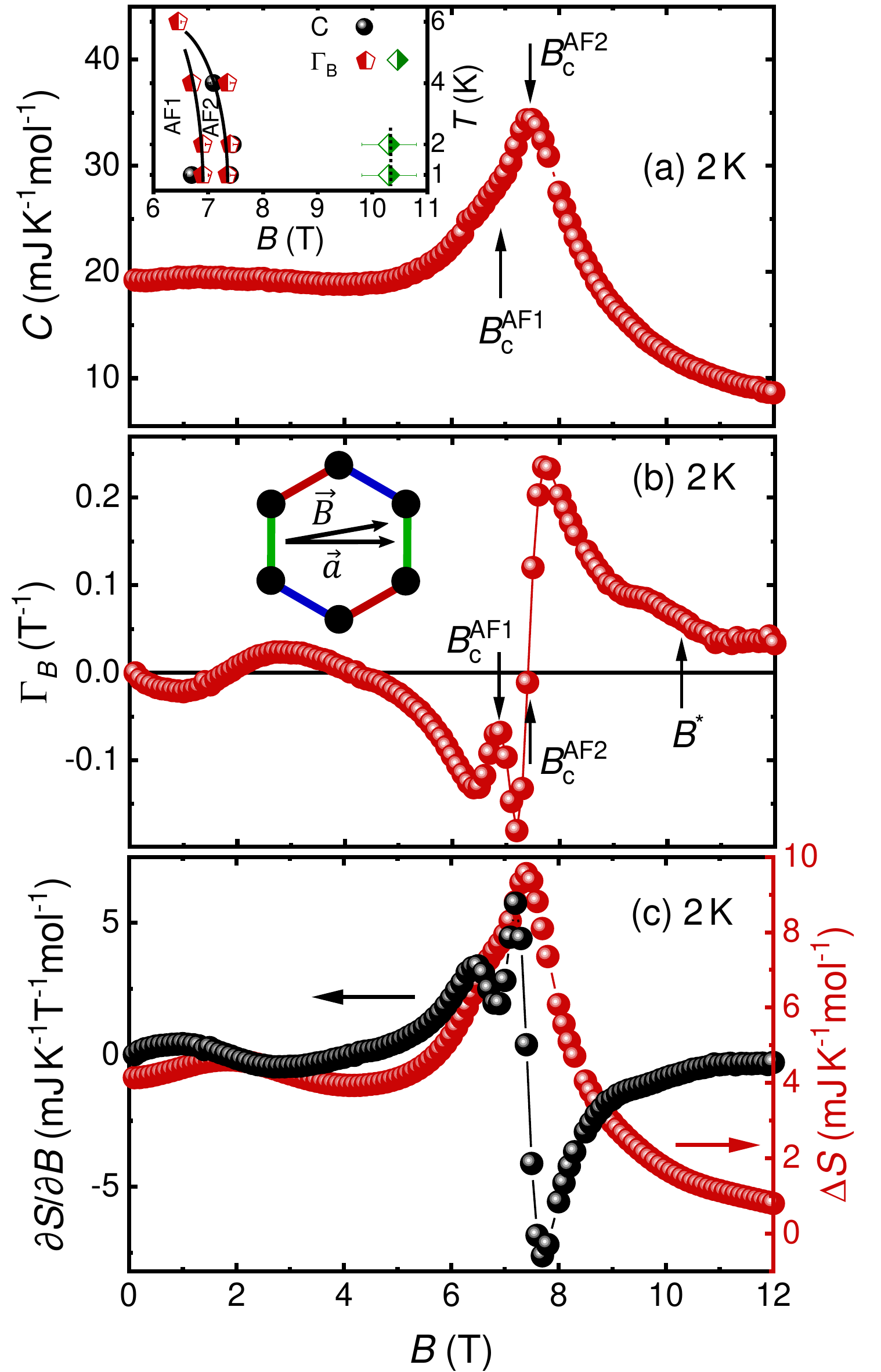}%
	\caption{Magnetic field dependencies of (a)
	the specific heat $C$, (b) the
	magnetic Gr\"uneisen parameter $\GammaMag$, (c) the field derivative of the entropy $\partial{}S/\partial{}B$ and entropy increment $\Delta S$ (see text) for $\alpha$-RuCl$_3$ at 2\,K. The inset in (a) shows a temperature-field phase diagram derived from our data, where solid lines stand for thermodynamic transitions and the dotted line for a crossover at $B^*$, and in (b) the field direction of our experiment.\label{fig.highFieldShoulder}\label{fig.phaseDiagram}}
\end{figure}

Here, we focus on $\alpha$-RuCl$_3$, which magnetically orders below 7\,K in zero field~\cite{Sears2015,Johnson2015} and reveals magnon excitations at low energies~\cite{Banerjee2017,Ran2017,wu2018field,Balz2019,wulferding2020magnon,sahasrabudhe2020high,Ponomaryov2020}. Additionally, it shows broad high-energy spectral features that are
often interpreted as fractionalized excitations -- vestiges of the proximate QSL
state~\cite{Sandilands2015,Nasu2016,Banerjee2017,ZheWang2017,Wellm2018} -- although this behavior can also be described in terms of magnon decays and incoherent excitations originating from strong magnetic anharmonicities
~\cite{Winter2017b,Winter2018,smit2020magnon,maksimov2020rethinking}.
In-plane magnetic fields lead to a gradual suppression of magnetic order that completely disappears around
$B_c^{\rm AF2}\simeq
7.5$\,T~\cite{Wolter2017,Sears2017,Wang2017,Leahy2017,Banerjee2018} (see also
the inset of Fig.~\ref{fig.phaseDiagram} for the data from our study). 

The nature of the phase lying immediately above $B_c^{\rm AF2}$ has been a
matter of significant debate. On the one hand,
it can be seen as a precursor of the
gapped fully polarized state~\cite{Baek2017}, but reveals only a fraction of
the total magnetization of \mbox{spin-$\frac12$} because of the sizable
off-diagonal exchange present in the system~\cite{Winter2018,Janssen2017}. On
the other hand, if magnetic order is seen as an obstacle to
the Kitaev QSL, then the suppression of the ordered phase should give way to the QSL
itself. This latter scenario was reinforced by the observation of quantized half-integer
thermal Hall effect, a signature of underlying topological
order~\cite{Nasu2017}. This quantization was initially reported at
$7-9$\,T~\cite{Kasahara2018}, right above $B_c^{\rm AF2}$, although more recent
studies detected quantized behavior only in higher in-plane fields of
$8.5-9$\,T~\cite{Yamashita2020} or even $10-12$\,T~\cite{Yokoi2020}, suggesting
that the putative spin-liquid phase may not emerge from the magnetically
ordered phase directly.  Importantly, if a topological QSL occurs at intermediate fields, the borders of the phase would have to be marked by distinct thermodynamic signatures~\cite{chen2010local}. 

In this Letter, we therefore examine the temperature-field phase diagram of
$\alpha$-RuCl$_3$ by high-resolution thermodynamic measurements and seek to
explore phase transitions related to the half-integer plateau in thermal Hall
effect of Refs.~\onlinecite{Kasahara2018,Yokoi2020,Yamashita2020}. We find that
no phase transitions occur for fields above $B_{c}^{\text{AF2}}$, whereas the 
previously reported shoulder anomalies in the magnetocaloric coefficient~\cite{Balz2019} and magnetostriction~\cite{Gass2020} above 8\,T are likely due to a change in the nature of the lowest excited states with only a
tiny change in entropy. This casts doubts on the existence of
an intrinsic phase transition
between the purported QSL and the partially-polarized 
phase of $\alpha$-RuCl$_3$.
We analyze the possible microscopic origin of the observed
shoulder anomaly via 
finite-temperature exact diagonalization for realistic spin models for $\alpha$-RuCl$_3$~\cite{rau2014generic,Winter2017b,kaib2020magnetoelastic}.
Our results suggest that this feature may originate from crossings of low-lying
excitations related to competing distinct phases, without the system
experiencing a phase transition to a new phase. 


Measurements of the magnetic Gr\"uneisen parameter ($\GammaMag$) and specific heat ($C$) as a function of field were performed with a dilution refrigerator using the high-resolution alternating-field method for $\GammaMag$~\cite{tokiwa-rsi11} and quasi-adiabatic heat pulse and relaxation method for $C$~\cite{Suppl}. High-quality single crystals were grown by vacuum sublimation~\cite{Reschke2018}. Sample quality was checked by a zero-field heat-capacity measurement. 
The sample showed a single phase transition around 7\,K and no signatures of an additional phase transition at 14\,K, which could be caused by stacking faults. The zero-field measurement repeated after the measurements in the magnetic field confirmed that the sample remained intact, with no stacking faults introduced during the experiment~\cite{Suppl}.

From previous works on $\alpha$-RuCl$_3$ it is known that the magnetic phase diagram varies somewhat for different in-plane field directions~\cite{lampenkelley2018}. For fields applied perpendicular to the Ru--Ru bonds (crystallographic $a$ direction), one observes an extended region of an intermediate ordered
phase~\cite{lampenkelley2018}, which is stable between $B_c^{\rm AF1}$ 
and $B_c^{\rm AF2}$, and presumably related to a change of the out-of-plane ordering wavevector~\cite{nagler}. 
The half-integer thermal Hall effect was observed for $B\,\|\,a$~\cite{Kasahara2018,Yokoi2020,Yamashita2020}, whereas from symmetry considerations no thermal Hall effect is expected for $B\,\|\,b$~\cite{Yokoi2020,gordon2020testing}. For our measurements, we choose a field direction $10^{\circ}$ away from $\vec{a}$ (Fig.~\ref{fig.phaseDiagram}b) following the setting of~\cite{Balz2019}.

In Fig.~\ref{fig.highFieldShoulder}, we show both specific heat $C$ and the magnetic Gr\"uneisen parameter $\GammaMag=-(\partial{}M/\partial{}T)/C=(1/T)(\partial{}T/\partial{}B)_S$, which quantifies the ratio between the temperature derivative of the magnetization and the specific heat. $\GammaMag$ is a measure of the adiabatic magnetocaloric effect and a very sensitive probe of classical and quantum phase transitions~\cite{ZhuUniversally2003,GarstM:SigctG,Tokiwa-Sr327}. Using the high-resolution alternating-field method~\cite{tokiwa-rsi11}, we determine magnetocaloric effect under perfect adiabatic conditions, in contrast to the previous magnetocaloric study of Ref.~\onlinecite{Balz2019}. 
Combining $\GammaMag$ with the specific heat provides access to the temperature derivative of the magnetic entropy across the phase diagram as $\partial{}S/\partial{}B=-C\,\GammaMag$.

In the specific heat, shown in Fig.~\ref{fig.highFieldShoulder}a, the dominant feature is a peak
at $B_c^{\rm AF2}=7.4\,$T. At the same field, $\GammaMag$ exhibits a sharp jump
with a sign change from negative to positive, cf.
Fig.~\ref{fig.highFieldShoulder}b. We note that entropy $S$ generally exhibits
a maximum at a second-order phase transition between the magnetically ordered
and paramagnetic phases~\cite{GarstM:SigctG}. The entropy change across the transition, $\Delta S=-\int dB\,C\,\GammaMag$ (Fig.~\ref{fig.highFieldShoulder}c), indeed shows a maximum, because $C$ is always positive, and a sign change of $\GammaMag$ from negative to positive with increasing field corresponds to a maximum in the entropy at $B_c^{\rm AF2}$.

Another anomaly at $B_c^{\rm AF1}=6.9\,$T is also clearly visible as a local
maximum of $\GammaMag$, although a corresponding feature in $C$ is nearly absent. For a weak first-order phase
transition one also expects a maximum of the entropy, but without a
discontinuity in $C$ if the transition is significantly smeared out. In this
case, $\partial{}S/\partial{}B$ goes through a minimum that causes a maximum in $\GammaMag$
without the jump and sign change. This way, we interpret $B_c^{\rm AF1}$ as a
first-order phase 
transition, which is compatible with the reported change in the
magnetic propagation vector at this field~\cite{nagler}. 
We note in passing that there is another sign change in $\GammaMag$ at $1.8\,$T, indicating an additional entropy maximum. It is mostly likely related to the domain reconstruction reported in previous studies~\cite{Sears2017}. 

Beyond $B_c^{\rm AF2}$, the antiferromagnetic (AF) order is destroyed. At higher fields, if a QSL phase exists, at least one additional phase transition is necessary when the QSL is suppressed, as 
indicated by the disappearance of the half quantization in the thermal Hall
effect~\cite{Kasahara2018}. However, we find 
no
signature of a further phase transition in our specific heat data (Fig.~\ref{fig.highFieldHeat}a). 
We find only a broad shoulder in $\GammaMag$ centered at $B^\ast\sim 10$\,T,
which is much weaker than the two other
anomalies. From these observations, we conclude that there is no second-order
phase transition above $B_c^{\rm AF2}$ within the resolution of our experiment.
As shown in the phase diagram (Fig.~\ref{fig.phaseDiagram}), the shoulder at
$B^\ast$ is observed also at 1\,K, but not above
2\,K~\cite{Suppl}. 
 According to recent Raman~\cite{wulferding2020magnon,sahasrabudhe2020high} and neutron-scattering~\cite{Balz2019} experiments, the field range of the shoulder falls into the region of a gapped phase.


To obtain further information on the shoulder-like anomaly at $B^\ast$ in the magnetic Gr\"uneisen parameter, we inspect the entropy contained in different transition anomalies using $\partial{}S/\partial{}B=-C\GammaMag$
shown in Fig.~\ref{fig.highFieldShoulder}c. Two clear jumps are observed at $B_c^{\rm AF1}$ and $B_c^{\rm AF2}$, whereas at $B^\ast$ only a broad
shoulder can be distinguished (Fig.~\ref{fig.highFieldHeat}b). Clearly, the
anomaly near $B^\ast$ results in a negative contribution to $\partial{}S/\partial{}B$ and
thus also to an additional decrease in the magnetic entropy. Qualitatively,
this may indicate that the state for  $B > B^\ast$  has lower entropy, which
naturally arises due to the polarization of moments by magnetic field. 

\begin{figure}[t!]
\includegraphics[width=0.45\textwidth]{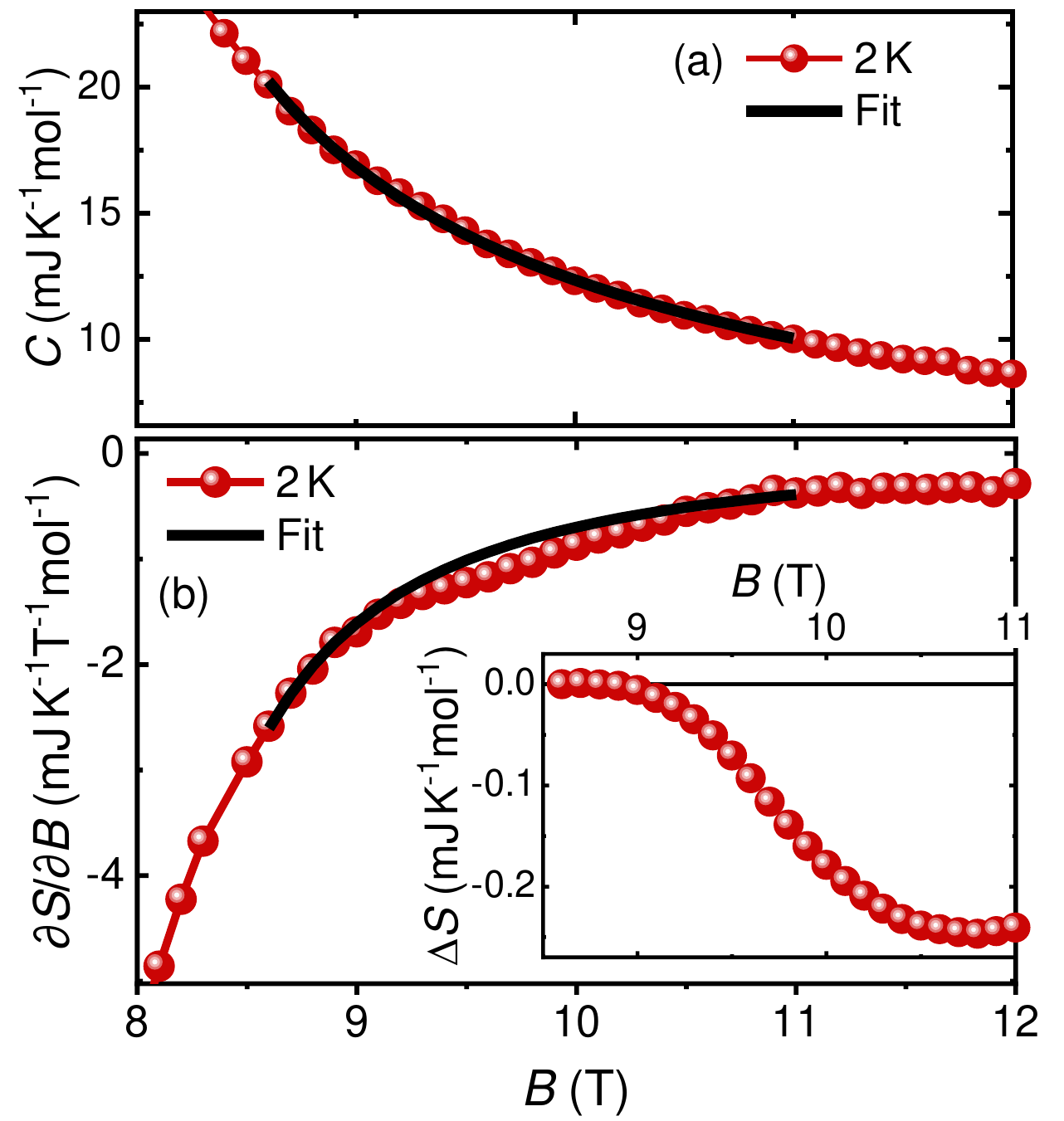}%
	\caption{Magnetic field dependence of the specific heat (a) and field derivative of the entropy (b), as well as the entropy change (inset) for $\alpha$-RuCl$_3$ at 2\,K between 8 and 12\,T. The black solid lines indicate $C=\beta(B-B_c)^\epsilon$ with $B_c^{\text{AF2}}=7.4$\,T,  $\beta=22.7(9)$mJK$^{-1}$mol$^{-1}$T$^{-\epsilon}$ and $\epsilon=0.64(5)$ (a) and $\partial{}S/\partial{}B=\beta(B-B_c)^\epsilon G_r/(B-B_c)$ (b), with $G_r=-0.157(2)$, respectively~\cite{Suppl}. The inset shows the difference between the measured entropy and the integration of the above function for $\partial{}S/\partial{}B$ for fields between 8.6 and 11\,T. \label{fig.highFieldHeat}}
\end{figure}

By subtracting the background~\cite{Suppl} and integrating $\partial{}S/\partial{}B$, we estimate
that only a tiny amount of entropy, 0.25\,mJ/mol$\cdot$K ($4.3\times 10^{-5}
R\log 2$), is associated with the shoulder at $B^\ast$. For comparison, the entropy
change at $B_c^{\rm AF2}$ is 12.5\,mJ/mol$\cdot$K and thus 50 times larger.
Although entropy changes generally become small at low temperatures, they are
not expected to become so tiny, especially at the transition between a
chiral Kitaev spin liquid and polarized state, where the flux gap is closed and low-energy
excitations are abundant~\cite{Nasu2014,kaib2019kitaev}. 

\begin{figure}[t!]
\includegraphics[width=0.45\textwidth]{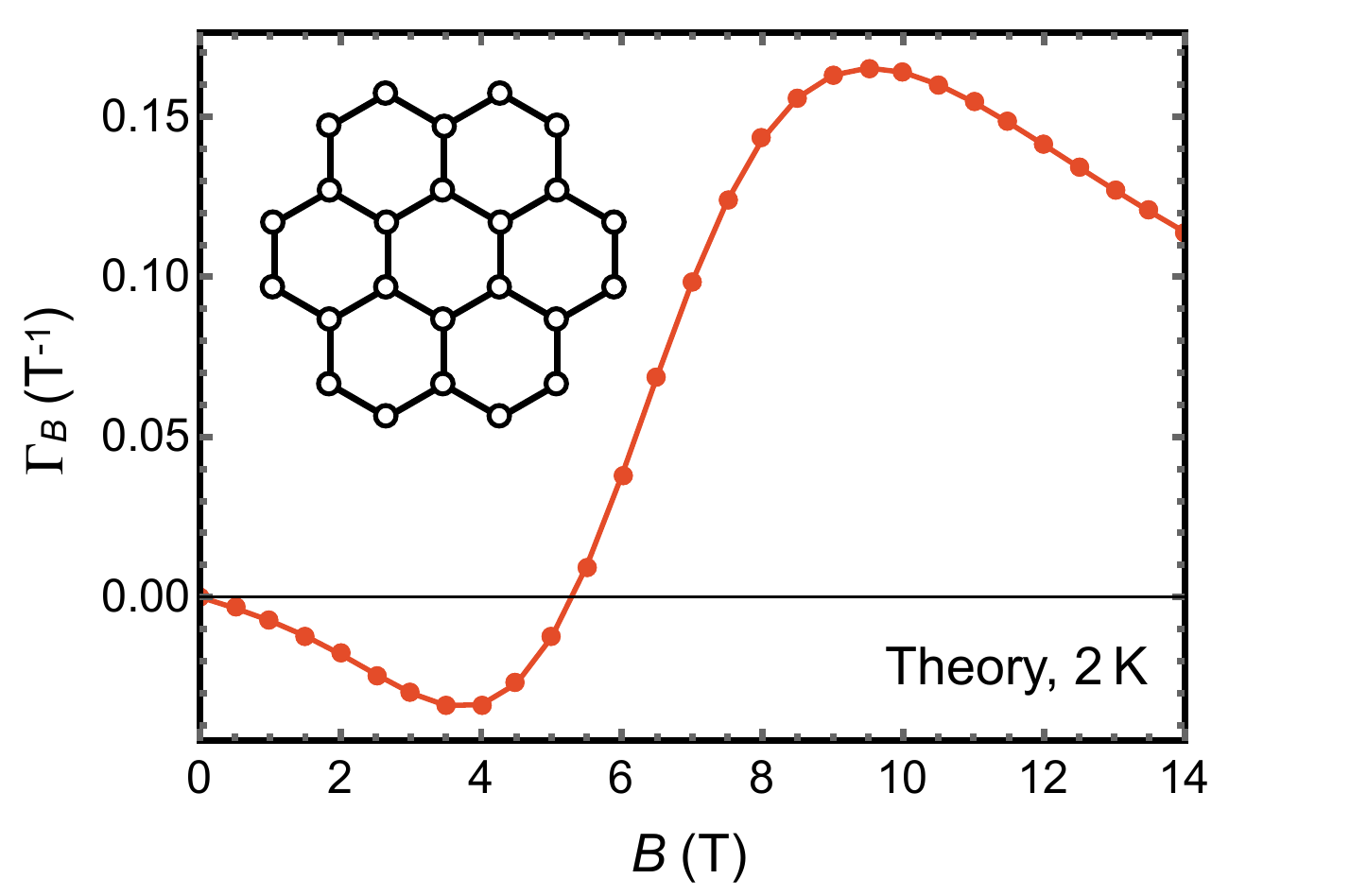}%
\caption{$T=2\,\text{K}$ exact diagonalization
	 results of the magnetic Gr\"uneisen parameter $\GammaMag$ for the minimal model of \mbox{$\alpha$-RuCl$_3$}~\cite{Winter2017b}. The inset shows the employed periodic cluster. 
\label{fig.ED.onepanel}}
\end{figure}

 To analyze our experimental results in the context of realistic Hamiltonians of $\alpha$-RuCl$_3$,  
we first perform exact diagonalization calculations on the two-dimensional \textit{ab-initio} guided minimal model of Ref.~\onlinecite{Winter2017b}, 
which reproduces various experimental aspects of \rucl~\cite{Winter2017b,Wolter2017,Winter2018,cookmeyer2018spin,riedl2019sawtooth} without hosting an in-plane
field-induced QSL phase. In particular, it reproduces the magnetic-field dependence of the magnetotropic coefficient~\cite{riedl2019sawtooth} recently reported~\cite{Modic2018,Modic2020}. We compute finite-temperature observables almost exactly~\cite{Suppl} 
on a two-dimensional 24-site periodic cluster shown in the inset of Fig.~\ref{fig.ED.onepanel}. We stress that these calculations cannot capture features related to three-dimensional effects (like $B_c^{\text{AF1}}$), and finite-size effects lead to a smearing out of phase transitions. Nonetheless, this model and method capture the essential field- and temperature evolution of the anomalies at $B_c^{\rm AF2}$ and of the overall magnitude for the Gr\"uneisen parameter (Fig.~\ref{fig.ED.onepanel}) and other measured quantities, as shown in the Supplemental Material~\cite{Suppl}. 
Focusing now on \GammaMag,
we observe that the computed absolute order of magnitude as well as the sign
change related to the suppression of zigzag order ($B_c^{\text{AF2}}$) at
$B\approx 6\,\text{T}$ in the model agree well with experiment. In the
partially-polarized phase of the model ($B>6\,\text T$),  \GammaMag\ reaches its
maximum not instantly at the phase transition, but at $B\approx 10\,\text{T}$,
which is likely related to the above mentioned finite-size effect. For all higher
field strengths, \GammaMag\ falls monotonically but stays positive. 


The results provided by this model \cite{Winter2017b}
 however lack a shoulder-like anomaly like the one observed experimentally at
 $B^\ast$. On the other hand,
 since this shoulder lacks the appearance of a
 phase transition \cite{ZhuUniversally2003}, we are led to ask, can anomalies occur in general
 Gr\"uneisen parameters $\Gamma_\lambda\equiv -(\partial S/\partial \lambda)/C$
 that are \textit{not accompanied by phase transitions}? The universal zero-temperature limit of
 $\Gamma_\lambda$ of all gapped phases is in fact markedly simple~\cite{Suppl}:
\begin{equation}
\Gamma_\lambda(T\rightarrow 0) 
= \frac{\Delta'}{\Delta}
  \label{eq.dgapbygap}
\end{equation}
where $\Delta$ is the gap between the ground state and lowest excited state and $ \Delta' \equiv \mathrm d \Delta /
\mathrm d \lambda$. Eq.~(\ref{eq.dgapbygap}) holds for both the magnetic Gr\"uneisen parameter
($\lambda=B$) measured in this study, as well as the structural one
($\lambda=$ pressure). 

\begin{figure}[t!]
\includegraphics[width=0.45\textwidth]{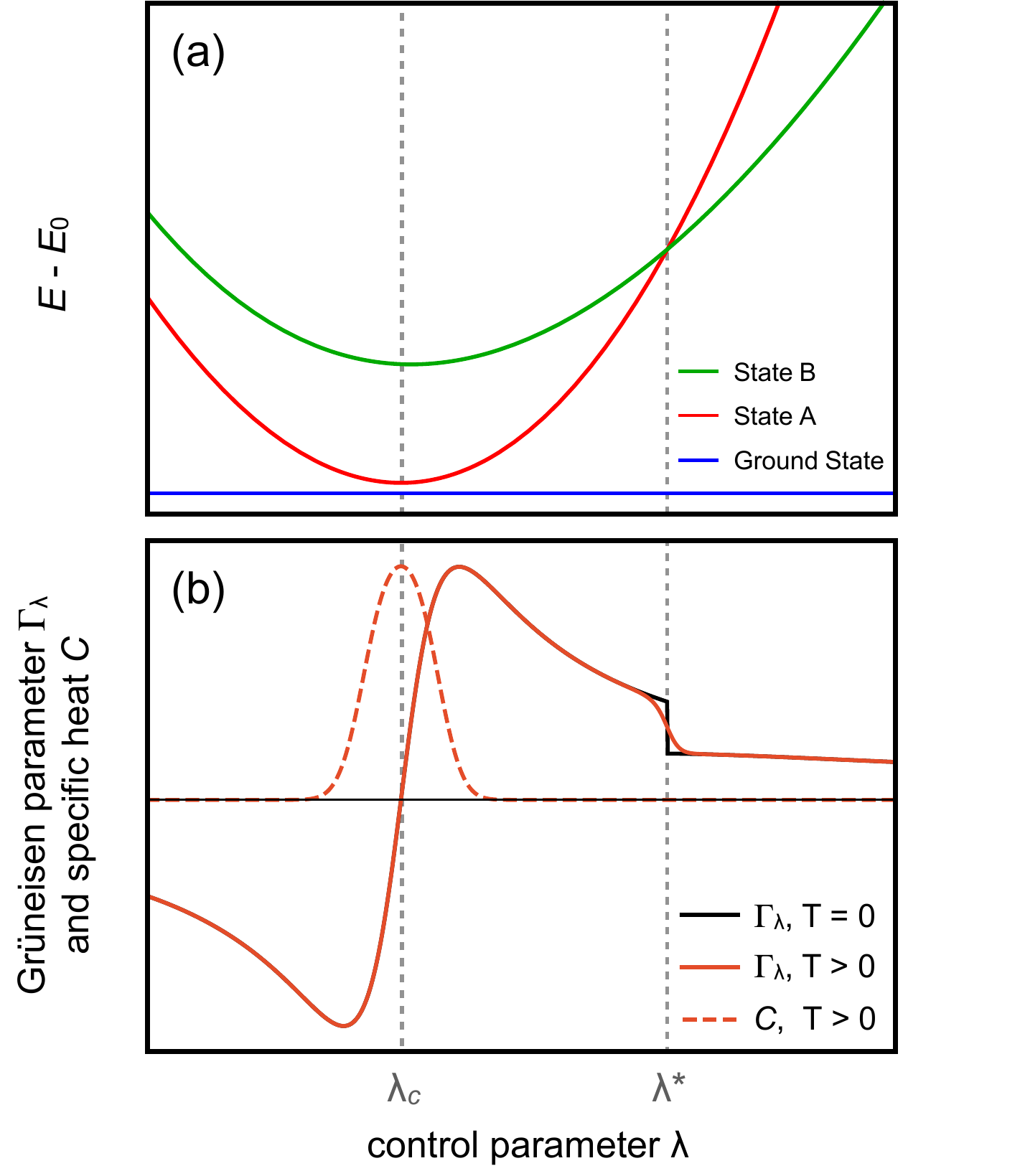}%
\caption{(a) Schematic of a discrete spectrum evolving with a control parameter $\lambda$. It contains an avoided level crossing at $\lambda_c$ and a level crossing in the lowest excited states at $\lambda^{\ast}$. These lead, respectively, to a sign change ($\lambda_c$) and a shoulder ($\lambda^\ast$) in the low-temperature Gr\"uneisen parameter $\Gamma_\lambda$~(b). If $\lambda_c$ is a quantum critical point, the gap would vanish at $\lambda_c$ in the thermodynamic limit, leading to a diverging $\Gamma_\lambda$ at $\lambda_c$ \cite{GarstM:SigctG,ZhuUniversally2003,Suppl}. 
The low-temperature specific heat $C$ [dashed curve] is nearly unaffected by the level crossing at $\lambda^\ast$. 
\label{fig.toyspectrum}
}
\end{figure}

From Eq.~(\ref{eq.dgapbygap}), we can anticipate two distinct types of anomalies in $\Gamma_\lambda$, which we illustrate via the schematic discrete spectrum shown in
Fig.~\ref{fig.toyspectrum} (note that Eq.~(\ref{eq.dgapbygap}) nevertheless
also holds for continuous spectra). 
If we consider a quantum phase transition  at a critical $\lambda_c$, marked by a gap closure \cite{sachdev_2011}, it is easy to see that $\Gamma_\lambda$ diverges \cite{Suppl} and changes sign from negative to positive upon closing ($ \Delta'<0$) and reopening ($\Delta'>0$) of the gap. Provided the gap closes or opens as a power law in the thermodynamic limit, $\Delta \propto |\lambda-\lambda_c|^p$, a general consequence of this formula is that $\Gamma_\lambda\propto (\lambda-\lambda_c)^{-1}$ regardless of the specific power~$p$. This recovers the known behavior for Gr\"uneisen parameters at quantum critical points \cite{ZhuUniversally2003,GarstM:SigctG}. 
However, anomalies may also occur due to level crossings between the
   lowest excited states, which are labelled $A$ and $B$ in Fig.~\ref{fig.toyspectrum}a. While this crossing is comparatively invisible to the low-temperature specific heat~$C$, the abrupt change in the slope of the gap ($\Delta'$) introduces a discontinuity in $\Gamma_\lambda$ without a divergence as shown in Fig.~\ref{fig.toyspectrum}b at $\lambda^\ast$. 
   At small finite temperatures, the drop in $\Gamma_\lambda$ is smeared out to a shoulder-like feature, that closely resembles $B^\ast$ in experiment. In our interpretation, $B^*$ therefore corresponds to an abrupt change in the nature of the lowest excitations, rather than to a phase transition.


Regarding \rucl, various scenarios may be consistent within this interpretation. The $B^*$ anomaly may occur when the $k$-point associated with the lowest energy excitations changes as a function of field within the partially-polarized phase. While this does not occur in the minimal
model~\cite{Winter2017b} corresponding to  Fig.~\ref{fig.ED.onepanel}, 
such an excited state level crossing is a recurring feature of models that are more proximate to a zero-field phase other than zigzag AF. For example, we have found \cite{Suppl} an anomaly in $\GammaMag$ for a more complete
 \textit{ab-initio} derived model \cite{kaib2020magnetoelastic} that is closer to ferromagnetic order at zero field, and includes
 additional interaction terms beyond those considered in the minimal model~\cite{Winter2017b}. In this case, the lowest-energy excitations switch from the zigzag wave vector to $\mathbf k = 0$ above a particular field strength within the gapped partially-polarized phase.  
Such a scenario can be verified via inelastic neutron scattering probes of the high-field dispersion of the magnetic excitations where a change or shift of the k-point with the softest mode may be observed around $B^\ast$~\cite{Suppl}. 
 In a more exotic scenario, the level crossings between the
 lowest excited states can also be induced by pushing the models closer to QSL
 phases, which we demonstrate by tuning nearer towards a hidden AF Kitaev
 point \cite{Suppl,kaib2019kitaev}. In both cases, our results imply the remarkable observation that the measured anomaly at $B^\ast$ may indicate that \rucl\  is proximate to competing phases at finite fields, but \textit{does not enter them}.   
 This poses the question whether the thermal Hall conductivity could also change anomalously at these field strengths without necessitating a phase transition, implying that no topological QSL would be entered or exited.

In summary, we have performed detailed high-resolution measurements of the specific heat and
magnetic Gr\"uneisen parameter of $\alpha$-RuCl$_3$ as a 
function of in-plane magnetic field. The observed two transitions at
$B_c^{\text{AF}1} = 6.6\,\text{T}$ and
$B_c^{\text{AF}2} = 7.4\,\text{T}$ are consistent with previous reports and correspond to
a transition between two AF states ($B_c^{\text{AF}1}$)
and to a transition from the second AF state
to the quantum paramagnetic state ($B_c^{\text{AF}2}$).
We also observe a third broad shoulder anomaly in $\GammaMag$ centered around
$B^\ast=10$\,T, consistent with previous studies~\cite{Balz2019,Gass2020}. 
This anomaly is invisible in the specific heat and inconsistent with a bulk phase
transition. Thus, the upper field limit of the claimed half-integer thermal Hall
plateau, which probably appears in this field range~\cite{Kasahara2018}, cannot be
explained by a phase transition between a presumed Kitaev QSL and polarized phase.  We instead propose an alternative origin of the high-field anomaly
as a change of the lowest-energy excitations without a phase transition, and
demonstrate numerically that this is compatible with realistic microscopic
models of $\alpha$-RuCl$_3$.

\begin{acknowledgments}
We acknowledge fruitful discussions with Steve Nagler, Kira Riedl, and Sananda Biswas. The work in Augsburg was supported by the German Research Foundation (DFG) via the Project No. 107745057 (TRR80) and by the Federal Ministry for Education and Research through the Sofja Kovalevskaya Award of Alexander von Humboldt Foundation (AAT). The work in Frankfurt was supported by the DFG Project No. 411289067 (VA117/15-1). 
\end{acknowledgments}

%

\clearpage\newpage
\begin{widetext}
\begin{center}
\large\textbf{\textit{Supplemental Material}\smallskip \\ Thermodynamic perspective on the field-induced behavior of $\alpha$-RuCl$_3$}
\end{center}
\end{widetext}

\renewcommand{\thefigure}{S\arabic{figure}}
\renewcommand{\thetable}{S\arabic{table}}
\renewcommand{\theequation}{S\arabic{equation}}
\setcounter{figure}{0}
\setcounter{equation}{0}

\begin{widetext}

\section{Selection of sample}
\aRuCl{} crystals may be subject to stacking faults that manifest themselves by the smearing out of the magnetic transition at 7\,K and the appearance of additional magnetic transitions at higher temperatures~\cite{Cao2016}. We checked the quality of our samples by measuring magnetization using MPMS 3 from Quantum Design. Three crystals were tested (Fig.~\ref{fig.sample_characterization}a). Crystals 1 and 2 show a weak bend around 14\,K that indicates the occurence of a second magnetic transition due to stacking faults. Crystal 3 does not show this bend. Zero-field specific heat of this crystal was further measured using Quantum Design PPMS both before and after dilution-refrigerator (DR) measurements of the specific heat and magnetic Gr\"uneisen parameter. The data in Fig.~\ref{fig.sample_characterization}b show the sharp $\lambda$-type anomaly at 7\,K with no signatures of additional transitions. This proves that neither our crystal of \aRuCl{} contained stacking faults prior to the measurements, nor were the stacking faults introduced during the DR measurement.

\section{Analysis of specific heat and magnetic \Gruneisen{} parameter}
Both specific heat and magnetic \Gruneisen{} parameter \GammaH{} are measured in our DR in the same setup, which sets the sample in quasi-adiabatic conditions with a weak thermal link to the bath~\cite{tokiwa2011, Li2018}. Both quantities can be measured in the same run without further warming up of the DR or exchanging the sample platform.

We measured specific heat by applying a short heat pulse of $\Delta t=1$\,s with a power of $\Delta{}P$ resulting in a fast increase of the sample temperature $\Delta{}T$, followed by a slower exponential relaxation with the characteristic time exponent $\tau$ (Fig. \ref{fig.analysis}a). For $\tau \gg\Delta t$, this temperature increase happens under quasi-adiabatic conditions, and therefore the specific heat is obtained by $C=\Delta Q/\Delta T$. In a real measurement a fitting of the exponential decrease is required in order to estimate $\Delta T$ (Fig. \ref{fig.analysis}b). If the exponential decrease is very fast in the case of a very small specific heat the determination of $\Delta T$ becomes challenging. Therefore the relaxation method has been used in this case \cite{Li2018}.\\

\begin{figure}[!hb]
\includegraphics[width=0.7\textwidth]{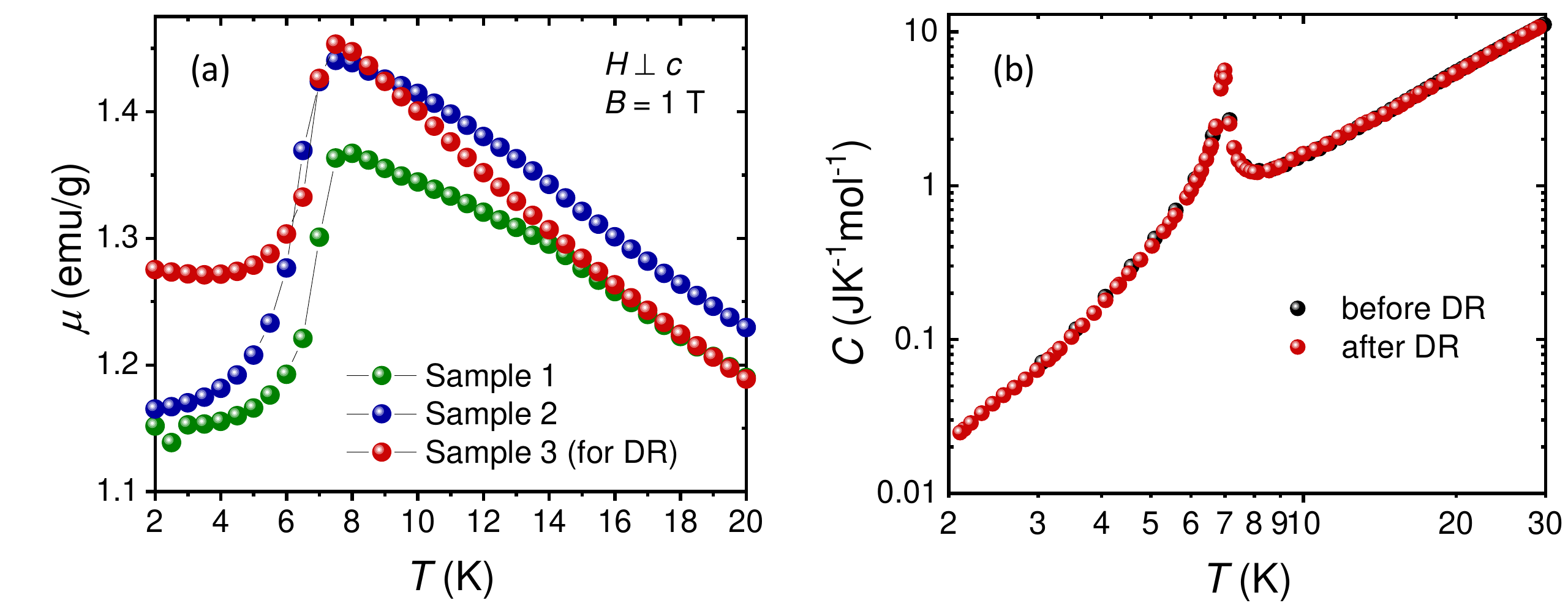}
\caption{(a) Temperature dependence of the magnetic moment $\mu$ for three crystals of $\alpha$-RuCl$_3$ measured at 1\,T for $H\perp c$. Sample 3 was used for the DR measurements. (b) Specific heat of sample 3 measured in zero field before and after the DR measurements. Both curves overlap nicely in the whole temperature range and show only one sharp peak at the ordering temperature $T=7$\,K. This confirms the absence of stacking faults in the crystal, both before and after the DR measurement.
\label{fig.sample_characterization}}
\end{figure}

\begin{figure}[t!]
\includegraphics[width=0.93\textwidth]{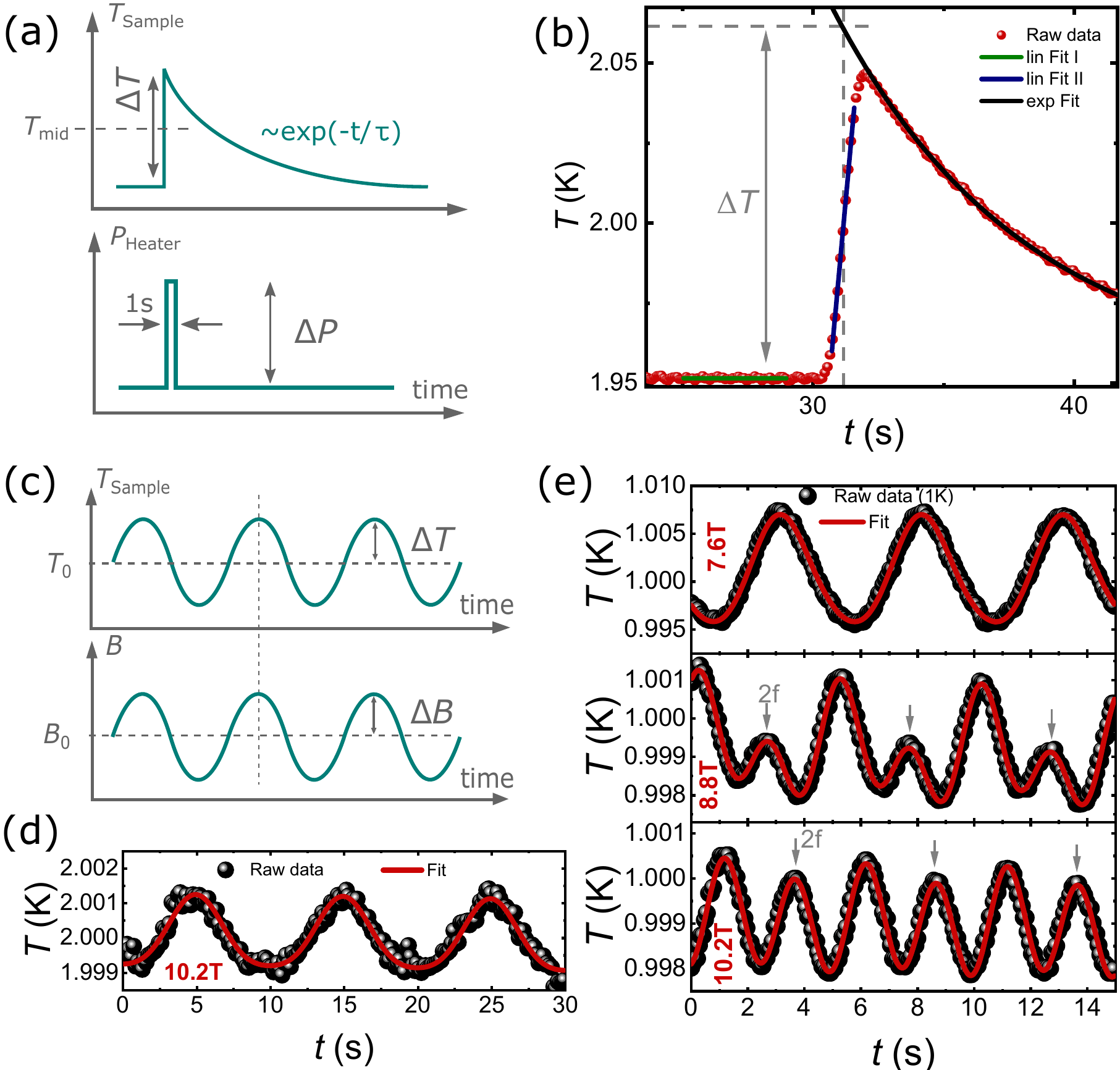}
\caption{(a) Illustration of the heat pulse method for the heat-capacity measurement. A power pulse $\Delta P$ applied for a short time $\Delta t=1$\,s creates a heat flow of $\Delta Q=\Delta P\cdot\Delta t$, which increases sample temperature by $\Delta T$. Afterwards the sample temperature decreases following a slow exponential decay with a characteristic relaxation time $\tau$. For $\tau \gg \Delta t$, the temperature increase $\Delta T$ occurs under quasi-adiabatic conditions and therefore the specific heat at $T_{\mathrm{mid}}$ is given by $C=\Delta Q/\Delta T$. (b) Example for the determination of $\Delta T$. Since the temperature increase is not a single step but rather has a finite slope, there is a small ambiguity in the determination of $\Delta T$. First, the starting temperature $T_{\mathrm{Start}}$ is determined by the linear fit I. Next, the exponential decay is fitted with $T_{\mathrm{exp}}(t)=T_i+A\cdot{}\exp(-t/\tau)$. Then $\Delta T$ is approximated by an equal area construction. (c) Principle of magnetocaloric (MCE) effect measurement. Additionally to the main magnetic field $B_0$, a small oscillating magnetic field is applied, $B_{\mathrm{ac}}(t)=\Delta B\cdot{}\sin{}(\omega t)$, leading to an oscillation of the temperature $T(t)=T_{\mathrm{mid}}+\Delta T\cdot{}\sin{}(\omega t+\phi)$. The magnetic \Gruneisen{} parameter is obtained as $\GammaH{}=1/T\cdot{}(\Delta T/\Delta B)$. (d) Raw data of $T(t)$ together with the fit at 2\,K and 10.2\,T ($f=0.1$\,Hz, $\Delta B=15$\,mT). No 2f contribution is visible. (e) Raw data of $T(t)$ together with the fit at 1\,K for three different fields. The frequency is $f=0.1$\,Hz and $\Delta B=15$\,mT. At 7.6\,T (upper panel), the signal of the sample is huge and can easily be fitted with the above mentioned formula. At higher fields, an additional feature with exactly twice the original frequency is clearly visible (grey arrows), therefore called the 2f contribution. This is typically due to eddy current heating in metallic parts of the cell. This contribution becomes prominent where the specific heat of the sample is very small, which is the case at high fields (compare the middle and lower panel with 8.8\,T and 10.2\,T, respectively). The raw data is fitted with Eq.~\eqref{eq.Gruneisen}.
\label{fig.analysis}}
\end{figure}

The magnetic \Gruneisen{} parameter is determined by measuring the magnetocaloric effect (MCE) of the sample. By applying an additional, oscillating magnetic field $B_{\mathrm{ac}}(t)=\Delta B\cdot{}\sin{}(\omega t)$ on top of the static field from the main magnet $B_0$, an oscillation of the sample temperature is induced with the same frequency $\omega =2\pi f$ and a phase shift $\phi$ (Fig. \ref{fig.analysis}c). For metals, a further oscillating contribution may occur due to eddy current heating with twice the original frequency, therefore called the 2f contribution. Overall, time-dependent sample temperature is fitted with~\cite{tokiwa2011} 
\begin{equation}
T(t)=T_{\mathrm{mid}}+\Delta T\cdot{}\sin{}(\omega t+\phi)+\Delta T_{\mathrm{2f}}\cdot{}\cos{}(2\omega t+\phi).
\label{eq.Gruneisen}
\end{equation}
Since \aRuCl{} is an insulator, the 2f contribution originates from metallic parts of the cell (e.g., from wires) and becomes dominant when specific heat of the sample is very small. That holds for high fields and low temperatures and can be seen in the 1\,K data of Fig.~\ref{fig.analysis}e. Even in the presence of the dominant 2f contribution, we can still extract the 1f contribution and determine the true MCE signal, as can be seen from the modulating temperature oscillation with the 1f frequency on the top of the dominant 2f oscillation. For temperatures of 2\,K (and above), no 2f contribution appears in high fields (Fig.~\ref{fig.analysis}d). Therefore, the shoulder-like feature at $B^\ast$ is certainly not an artefact due to the 2f fitting.
\section{Background subtraction}
At low temperatures, thermodynamic response of \aRuCl{} may be weak and obscured by the background contribution. Here, we explain the background subtraction procedure. 
\subsection{Specific heat}
In our setup, the sample is mounted with a small amount of Apiezon N-Grease. In the same way, the thermometer is fixed on top of the sample. This N-Grease as well as the thermometer, heater, and the rest of the cell contribute as an addenda \Cplatform{} to the measured absolute value of the specific heat \Ctot{}. In two separate measurement runs, we determine first the specific heat \Ctot{} of the sample together with the cell, and in a second step \Cplatform{} of the cell without the sample. The specific heat of the sample is obtained by subtracting the background, $\Csample=\Ctot-\Cplatform$ (Fig \ref{fig.backgroundHC}b).

\begin{figure}[t!]
\includegraphics[width=0.9\textwidth]{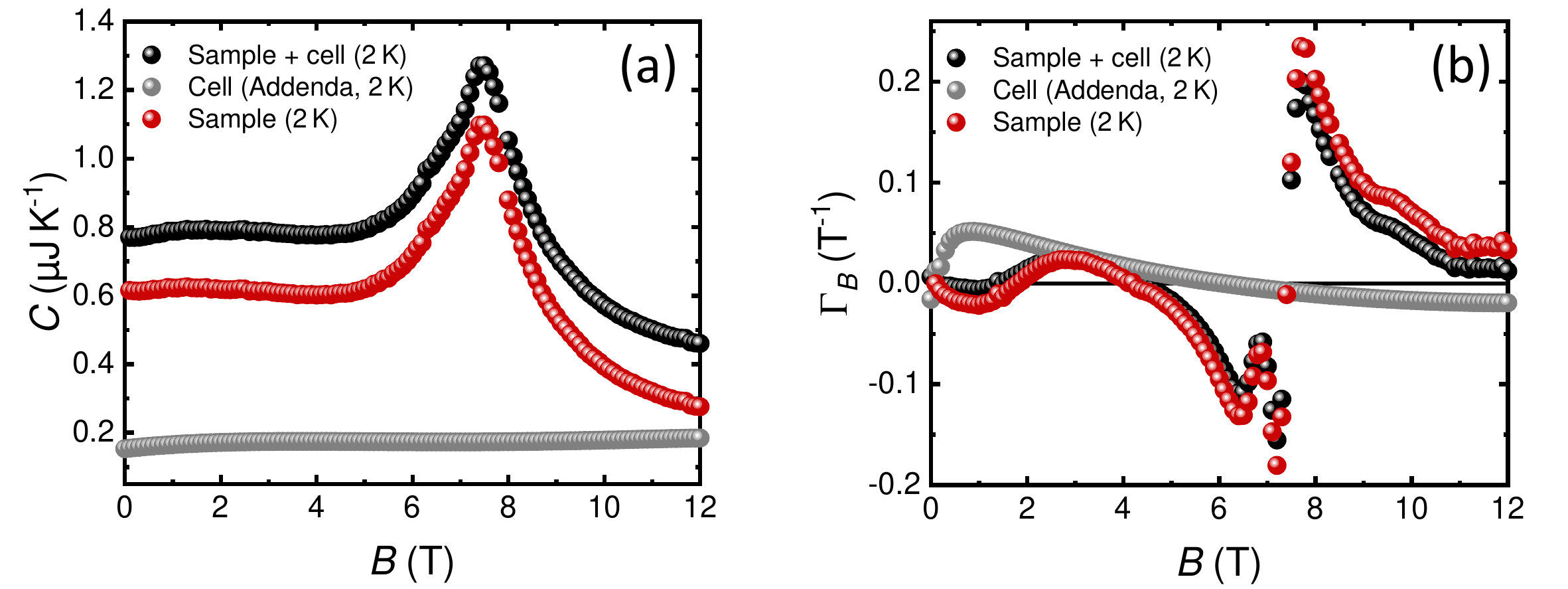}
\caption{(a) Measured absolute values of the specific heat for the whole setup (black) and cell only (grey). The specific heat of the sample (red) is obtained by the subtraction, $\Csample=\Ctot-\Cplatform$. (b) In the case of the magnetic \Gruneisen{} parameter, the contribution of the cell may be non-monotonic. The subtraction procedure following Eq.~\eqref{eq.GammaH_Background} is used to calculate the \Gruneisen{} parameter of the sample \GammaH[,Sa] (red), as described in the text.
\label{fig.backgroundHC}}
\end{figure}
\subsection{\Gruneisen{} parameter}
Also the magnetic \Gruneisen{} parameter \GammaH{} is affected by the addenda, because its specific heat is comparable to that of the sample. Here, the situation is more complicated, since \GammaH{} depends not only on specific heat, but also on the temperature derivative of the magnetization, both having background contributions from the cell:
\begin{equation}
\GammaH[,tot]=-\frac{\dMdT[tot]}{\Ctot}=-\frac{\dMdT[Sa]+\dMdT[Cell]}{\Csample+\Cplatform}=\frac{\GammaH[,Sa]\cdot\Csample+\GammaH[,Cell]\cdot\Cplatform}{\Csample+\Cplatform}
\end{equation}
This can be transformed into the following form:
\begin{equation}
\GammaH[,Sa]=\GammaH[,tot]+\frac{\Cplatform}{\Csample}\left(\GammaH[,tot]-\GammaH[,Cell]\right).
\label{eq.GammaH_Background}
\end{equation}
At $\Csample\gg\Cplatform$, the cell does not affect \GammaH[,Sa]. But as soon as $\Csample\sim\Cplatform$ like in the case of our \aRuCl{} sample at very low temperatures, two separate measurements with and without the sample are required again. In combination with the specific heat data, the background can be subtracted using Eq.~\eqref{eq.GammaH_Background}. The background \GammaH[,Cell] has been measured in the same run as the specific heat background \Cplatform.
\section{Estimation of the entropy related to \Bc[1] and \Bc[3]}
\begin{figure}[t!]
\includegraphics[width=0.85\textwidth]{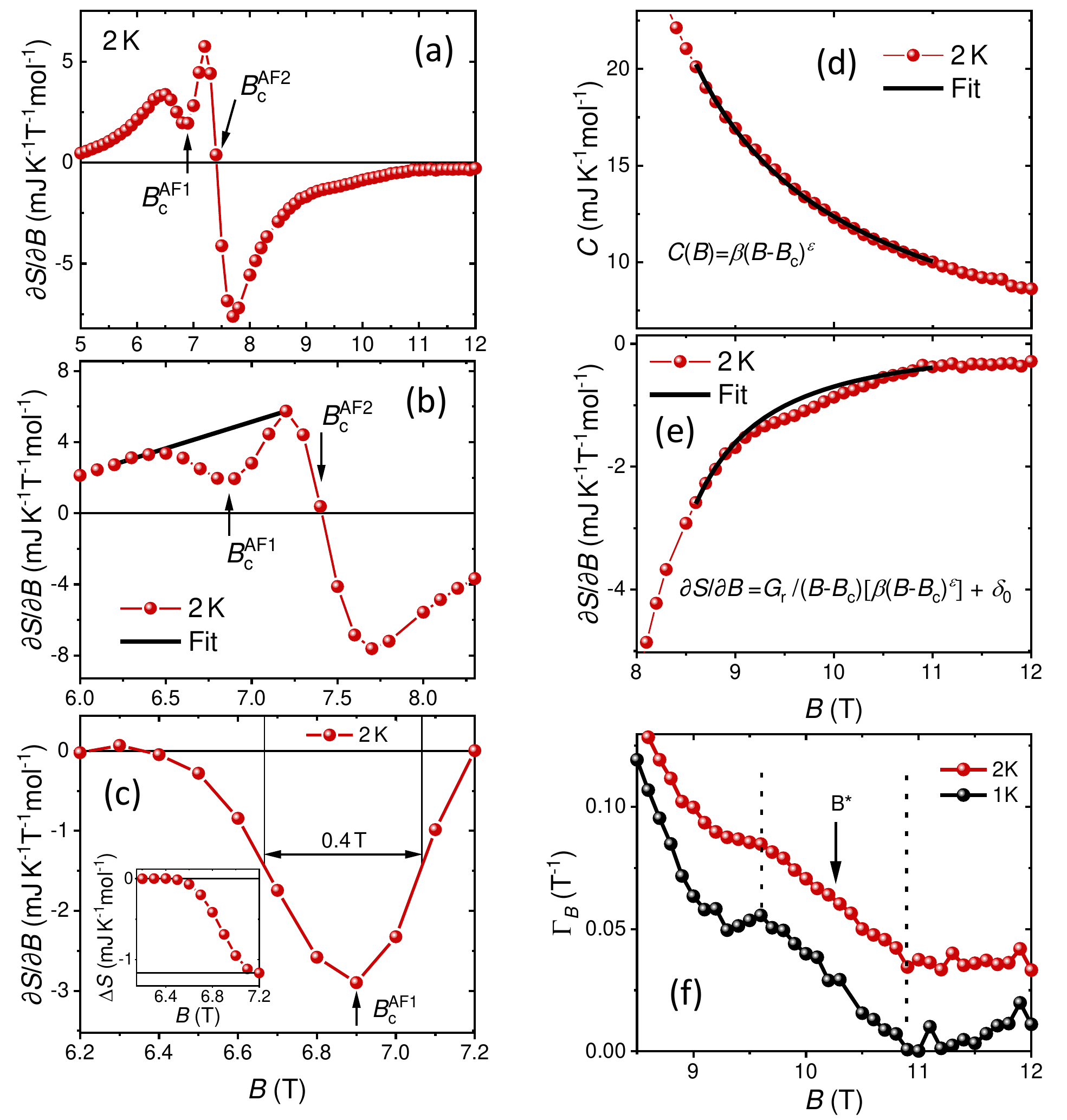}%
\caption{(a) Magnetic field dependence of the entropy derivative \dSdB{} including the anomalies \Bc[1], \Bc[2] and \Bc[3]. (b) Zoom into the field derivative \dSdB{} at \Bc[1] together with the "background" estimation of the dominant transition at \Bc[2] using a linear fit (data points of \Bc[1] neglected). (c) After subtraction of the fit, the \Bc[1] anomaly shows a minimum with a FWHM of $\sim 0.4$\,T. By integrating this minimum, we obtain field dependence of the relative entropy. The inset depicts the entropy step of $\sim 1.2(1)$\,mJK$^{-1}$mol$^{-1}$ associated with \Bc[1]. (d) Empirical fit of the field-dependent specific heat with fixed $B_c=7.4$\,T, resulting in $\epsilon=-0.64(5)$ and $\beta$=22.7(9)\,mJ\,K$^{-1}$mol$^{-1}$T$^{-\epsilon}$. (e) By using the parameter from the specific heat as fixed values we were able to fit the "background" due to the \Bc[2] transition, utilizing $\dSdB{}=-\GammaH C$ as explained in the text with fit results of$G_\mathrm{r}=-0.157(2)$ and $\delta_0=0.05(1)$\,mJ\,K$^{-1}$T$^{-1}$mol$^{-1}$. After subtraction of the fit only the contribution of \Bc[3] remains. By integration we calculated the related entropy and a step of $\sim 0.25(5)$\,mJK$^{-1}$mol$^{-1}$ for \Bc[3] which is shown in the main text of this Letter. (f) Determination of the position of \Bc[3]. \GammaH{} exhibits a sharp step in the zero temperature limit in the case of level crossing of the first excited states (see main text and Fig. \ref{fig:ShoulderModels}b,d). At finite temperatures this step is smeared out and we identify \Bc[3] as the midpoint between start and end of this regime (see dotted lines).
\label{fig.highFieldHeat}}
\end{figure}

Above the phase transition at $\Bc[2]=7.4$\,T, specific heat at 2\,K does not show any visible anomaly (cf. Fig.~\ref{fig.highFieldHeat}d, taken from the main text). Therefore, a thermodynamic 2$^{\mathrm{nd}}$ order phase transition can be excluded. However, a weak anomaly in the \Gruneisen{} parameter appears, denoted as \Bc[3] in the main text. Consequently, $\dSdB{}=-\GammaH{}C$ shows an additional negative contribution at \Bc[3], too (Fig.~\ref{fig.highFieldHeat}a). A similar behavior can be seen at \Bc[1], but here resulting in a clearly visible minimum. Much more entropy is involved in this anomaly, yet the behavior at \Bc[1] and especially at \Bc[3] is strongly influenced by the dominant contribution from \Bc[2]. Thus a direct determination of the related entropies is not possible. In the following, we explain the subtraction procedure utilized to estimate the entropy changes associated with \Bc[1] and \Bc[3], respectively, by considering two different approaches to the ``background'' due to \Bc[2].

First, we look at \Bc[1] at 2\,K and estimate the ``background'' due to \Bc[2] by a linear fit, where the data points related to \Bc[1] have been ignored (Fig.~\ref{fig.highFieldHeat}b). After subtracting the linear fit, a negative peak at \Bc[1] remains with the Full Width at Half Maximum (FWHM) of $\sim 0.4$\,T and an entropy step of 1.2\,mJ\,K$^{-1}$mol$^{-1}$ (Fig.~\ref{fig.highFieldHeat}c).

Since the anomaly \Bc[3] is much weaker, a simple linear fit of the ``background'' would not be adequate. Therefore, we used a special fitting function, assuming critical fluctuations due to the dominant second-order transition at \Bc[2]. First, $C(B)$ is fitted in the range of 8.6\,T to 11\,T with an empirical function $C=\beta(B-B_c)^\epsilon$ with a fixed value of $B_c=7.4$\,T, resulting in $\epsilon=-0.64(5)$ and $\beta=22.7(9)$\,mJ\,K$^{-1}$mol$^{-1}$T$^{-\epsilon}$. This fit is displayed as the solid line in Fig.~\ref{fig.highFieldHeat}d. For the magnetic Gr\"uneisen parameter $\GammaH{}$, a $G_r/(B-B_c)$ dependence has been predicted in the vicinity of a quantum critical point~\cite{garst2005}. We use this to obtain a formula for the ``background'' with $\dSdB{}=-C\GammaH=G_r/(B-B_c)[\beta(B-B_c)^\epsilon]+\delta_0$ , with fixed values from the specific heat fit, and $G_\mathrm{r}$ and an offset $\delta_0$ being the only fit parameter. We fit $\dSdB{}$ for 8.6\,T$\leq B\leq$11\,T by excluding the field range of the shoulder anomaly. It results in $G_\mathrm{r}=-0.157(2)$ and a very small offset $\delta_0=0.05(1)$\,mJ\,K$^{-1}$T$^{-1}$mol$^{-1}$ giving rise to the solid line in Fig.~\ref{fig.highFieldHeat}e.  The subtraction of this fit and integration of the difference leads to the step in the entropy mentioned in the main text being roughly a factor 5 smaller compared to \Bc[1]. A summary for the same analysis procedure at 1\,K and 4\,K (only estimation for \Bc[1] possible) can be found in Table \ref{tab.Bc1_Bc3}.

\begin{table}
\caption{Summary table for the estimations of the entropy related to \Bc[1] and \Bc[3] calculated by subtracting the contribution in \dSdB{} of the dominating \Bc[2] by a fit. Further details are explained in the text.}
\begin{ruledtabular}
\begin{tabular}{ccccccc}
\textit{T}(K)&\Bc[1](T)&FWHM$_1$(T)&$\Delta{}S_{1}$(JK$^{-1}$mol$^{-1}$) &\Bc[3](T)&$\Delta{}S_3$(JK$^{-1}$mol$^{-1}$) \\
\hline
1&6.9(1)&0.3(1)&0.20(5) &10.3(5)&0.026(5) \\
2&6.90(5)&0.40(5)&1.2(1) &10.3(5)&0.25(5) \\
4&6.7(1)&0.3(1)&1.40(3) &-&- \\

\end{tabular}
\end{ruledtabular}

\label{tab.Bc1_Bc3}
\end{table}

\section{Further results from 1\,K up to 6\,K}
\begin{figure}[t!]
\includegraphics[width=0.9\textwidth]{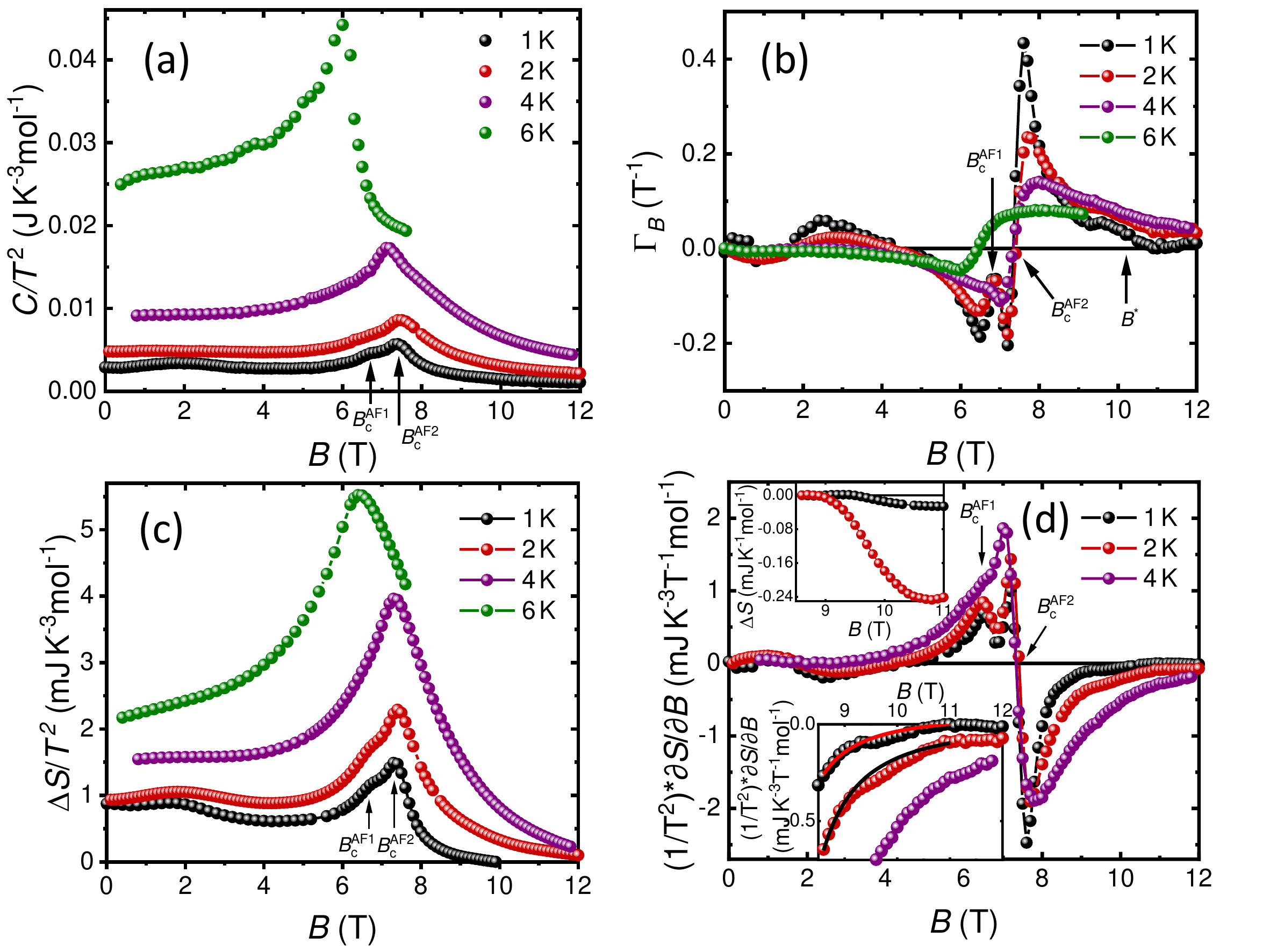}
\caption{Field dependence at 1\,K, 2\,K, 4\,K, and 6\,K. (a) Specific heat. For a better comparison $C/T^2$ has been plotted here. The cell addenda was measured only up to 5\,K therefore no addenda has been subtracted for 6\,K. Since the field-dependence of the background is rather weak at $T>5$\,K we assume that the addenda is nearly constant over the whole field range at this temperature and thus only contributes as a constant offset. All temperatures show the peak assigned to the transition out of the AF order. This peak marks \Bc[2] and shifts toward lower magnetic fields with increasing temperature. Only at the lowest temperatures of 1\,K and 2\,K the phase transition at $\Bc[1]<\Bc[2]$  appears as an additional weak anomaly. At fields above \Bc[2], no anomaly is visible. (b) Magnetic \Gruneisen{} parameter \GammaH{}. For all the temperatures, \GammaH{} shows a clear sign of a second-order phase transition at \Bc[2], namely, a sharp step with a sign change. While the 1\,K and 2\,K data clearly show a feature at \Bc[1], it becomes nearly indistinguishable at 4\,K. The weak shoulder at \Bc[3] is seen at 1\,K and 2\,K only. (c) Entropy increment $\Delta S$ plotted as $\Delta S/T^2$ for a better comparison. Only the main transition at \Bc[2] is visible. (d) Magnetic field derivative of the entropy scaled again by $T^2$. Similar to the 2\,K data from the main text (shown for comparison), there appears an extremely weak anomaly at \Bc[3] at 1\,K, shown in the lower inset. The relative entropy change is summarized in Table \ref{tab.Bc1_Bc3} and in the upper inset. No features associated with $B^\ast$ are seen above 2\,K.
\label{fig.furtherResults}}
\end{figure}

The specific heat and entropy of \aRuCl{} decrease rapidly towards low temperatures~\cite{Do2017}. Therefore, in Fig.~\ref{fig.furtherResults} we scaled the specific heat, entropy, and magnetic field derivative of the entropy by $T^2$ for a better comparison.

The main peak of the specific heat at \Bc[2] shifts to lower fields with increasing temperature. Only a very weak kink is visible at \Bc[1], most prominent at the lowest temperature of 1\,K. No anomaly is present above \Bc[2]. This is very similar to the behavior of the entropy increment $\Delta S$.

The magnetic \Gruneisen{} parameter (Fig.~\ref{fig.furtherResults}b) also shows a strong temperature evolution. The 2$^{nd}$ order phase transition at \Bc[2] shifts in accordance with the specific heat toward lower fields at higher temperatures. The additional, clearly visible feature at \Bc[1] is strongly smeared out at 4\,K and completely vanishes at 6\,K. The shoulder-like feature at \Bc[3] is only visible for 1\,K and 2\,K.

All three anomalies are also present in a very similar way in the field derivative d$S/$d$B$. For the 1\,K data, we fitted the high-field shoulder at \Bc[3] and used this fit for an approximation of the related entropy in the same way like for 2\,K. Compared to the maximum value of the peak at \Bc[2], the shoulder contributes 1.8\,\% of the entropy, which is comparable to the value at 2\,K.

\section{Generalized Kitaev models: Calculation details and further results}
We focus on generalized honeycomb Kitaev models under in-plane magnetic fields. Under the assumption of $C_3$ symmetry of the lattice and local $C_{2h}$ symmetry of the bonds, the Hamiltonian can generally be written as~\cite{rau2014}
\begin{align}
H = \sum_{ij} \mathbf{S}_i \cdot \mathbb{J}_{ij}\cdot  \mathbf{S}_j - g_{ab}\mu_B  \sum_i \mathbf{B} \cdot \mathbf{S}_i,\quad \text{where } \mathbb{J}_{ij} = \left(
 \begin{array}{c|ccc} 
 &\alpha&\beta&\gamma\\
 \hline
 \alpha & J_n & \Gamma_n  & \Gamma_n^\prime  \\
 \beta & \Gamma_n & J_n & \Gamma_n^\prime  \\
 \gamma & \Gamma_n^\prime& \Gamma_n^\prime & J_n+K_n
 \end{array}\right).
\end{align}
The bond-dependency of generalized Kitaev models is encoded in the variables $n,\alpha,\beta,\gamma$ in $\mathbb J_{ij}$: The subscript $n$ refers to $n$-th neighbor bonds and the directional dependence is given through $(\gamma,\alpha,\beta)=(x,y,z)$ 
for X-bonds, $(\gamma,\alpha,\beta)=(y,z,x)$ for Y-bonds, and $(\gamma,\alpha,\beta)=(z,x,y)$ for Z-bonds. In the literature, depending on the model, up to third-nearest neighbor bonds are expected to be relevant, i.e.~$n\le 3$. Then, in the notation of the $C2/m$ crystal structure, $n=1$ and $n=3$ Z-bonds are parallel to the crystallographic $b$ axis, while $n=2$ Z-bonds are perpendicular to $b$. The respective X-bonds (Y-bonds) are those with a $60^\circ$ ($120^\circ$) angle to the Z-bond. 

If $K_1$ in $\mathbb J_{ij}$ is the only nonzero coupling, the model reduces to pure Kitaev model, which is exactly solvable for $B\rightarrow 0$. 
For our main comparison to experiment we discuss the \textit{ab-initio} guided model for \aRuCl\ of Ref.~\onlinecite{Winter2017bs}, where the nonzero interactions are $(J_1,K_1,\Gamma_1,J_3)=(-0.5,-5,2.5,0.5)\,$meV, and $g_{ab}=2.3$~\cite{Winter2018s}. We refer to this here as the \textit{minimal model}. 

In order to solve the generalized Kitaev model, we resort to exact diagonalization techniques on finite clusters. Established methods for calculations at $T>0$ like the finite-temperaure Lanczos method (FTLM) \cite{jaklivc1994lanczos} or thermal pure quantum states \cite{sugiura2012thermal} often rely on statistical sampling of the Hilbert space. This bears large statistical errors at low temperatures. Accordingly our attempts of using FTLM for the present study failed to reach anywhere close to convergence after $R=500$ random starting vectors for temperatures $T\lesssim 3\,$K. We therefore instead apply a simpler method where we replace the statistical error (growing with decreasing $T$) by a systematic error (growing with increasing $T$) as follows: We calculate the $d_c$ lowest-energy eigenstates $\ket n$ of the Hamiltonian numerically exactly~\cite{lehoucq1998arpack} and approximate the canonical sums by restricting them to these low-energy eigenstates:
 \begin{equation}
  Z\approx \sum_{n=0}^{d_c-1} e^{-E_n/(k_BT)}, \quad \braket{O}= \frac1Z  \sum_{n=0}^{d_c-1} e^{- E_n/(k_B T)} \braket{n|O|n}.
\end{equation}
One may expect this approximation to be reasonable for $k_BT \ll (E_{d_c-1}-E_0)$. We show in \cref{fig:POdetails}a the obtained lowest $d_c=16$ eigenstates as a function of field strength for the model we mainly focus on. In this case $(E_{d_c-1}-E_0)/k_B\simeq 26\,$K at the field strength where it is the lowest. For a more thorough estimate of the cutoff error, we track $\braket O$ as a function of $d_c$ for different temperatures and only work further with results at temperatures where $\braket O$ is sufficiently converged with respect to the maximum $d_c$ we employ. Examples are shown in \cref{fig:POdetails}b,c for the magnetization $\braket M$ and the 
entropy $S=k_B \log Z + \frac{\braket H}{T}$
at $B=6$\,T, i.e.~at the critical field of the model, where the systematic error is the worst. Following such analysis, we have good confidence in results for temperatures below $T\leq 2.5\,$K for the minimal model with the employed cutoff $d_c=16$. 

\begin{figure}[!b]
	\includegraphics[width=1.0\linewidth]{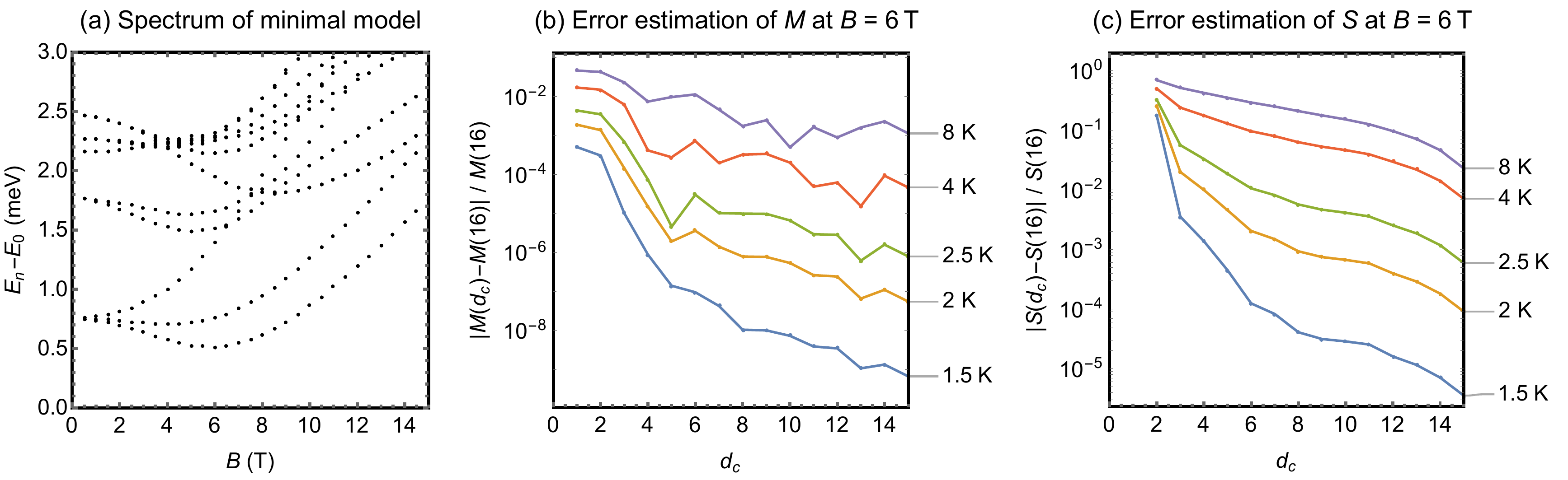}
	\caption{
	Exact diagonalization data for the minimal model \cite{Winter2017bs}, obtained on the 24-site cluster shown in Fig.~3  of the main text. (a)~Low-energy spectrum as a function of field strength. (b,c) Relative changes in the results for magnetization~(b) and entropy~(c) at $B=6\,$T as functions of the employed cutoff $d_c$, see \cref{eq:CutoffFormula}.
	\label{fig:POdetails}}
\end{figure}

The $T=2\,$K \Gruneisen\ parameter of the minimal model was shown in Fig.~3 of the main text. Further results are condensed in \cref{fig:POResults}, which shows the same quantities as in \cref{fig.furtherResults} at temperatures where we are confident about the systematic error introduced by the cutoff. As the calculations are performed on a two-dimensional 24-site honeycomb cluster (shown in Fig.~3 of the main text), we can not observe the \Bc[1] transition related to a change in the inter-plane order, and generally suspect a ``washing out" of thermodynamic phase transitions. Accordingly, we observe broad anomalies in the calculated quantities at the transition between the zigzag and partially-polarized phases (\Bc[2]). These manifest in maxima in the specific heat $C$ and entropy $S$ [\cref{fig:POResults}(a,c)] and sign changes in $\GammaH$ and $\partial S/ \partial B$ [\cref{fig:POResults}(b,d)] in accordance with experiment and as expected for such a phase transition. 
The shifting of \Bc[2] to lower fields with increasing temperature is also reproduced in this model, best seen by tracking the field where $\GammaH$ changes sign in \cref{fig:POResults}b for different temperatures. We note however that compared to experiment this shift appears to happen already at lower temperatures, which may be related to finite-size effects or the incompleteness of the minimal model. We note that three-dimensional couplings are expected to further stabilize the AF ordered phases~\cite{janssen2020magnon}. Aside from the position and qualitative temperature-evolution of the anomalies related to \Bc[2], the calculations also capture the approximate absolute orders of magnitudes of the measured observables outside of the critical field. 
\begin{figure}
	\includegraphics[width=0.8\linewidth]{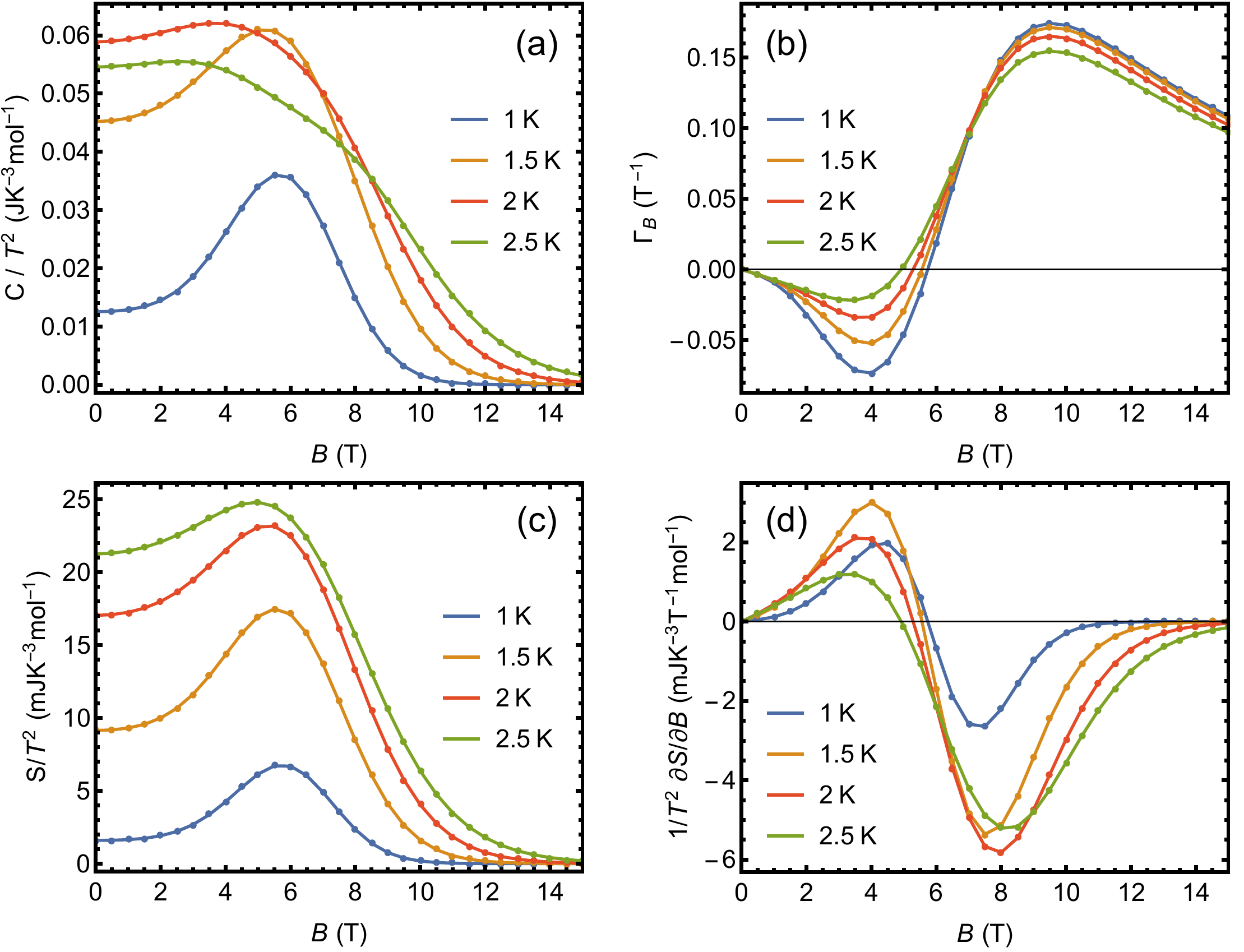}
	\caption{Further exact diagonalization results for the minimal model \cite{Winter2017bs} for the same observables as in \cref{fig.furtherResults}, with the exception of (c), which shows absolute entropy. \label{fig:POResults}}
\end{figure}

Missing in the minimal model is however the shoulder-like anomaly at \Bc[3], which is strongest seen in the \Gruneisen\ parameter, see \cref{fig.furtherResults}(b). As argued in the main text, we interpret this anomaly as an abrupt change in the nature of the lowest-energy excitations, which translates to a level crossing between the lowest excited states for a discrete spectrum. No such excitation level crossing takes place for any $B$ in our calculations for the minimal model, see \cref{fig:POdetails}(a). 
To demonstrate that such a feature above \Bc[2] is nevertheless realistic in generalized Kitaev models, we first consider the fully-\textit{ab-initio} model of Ref.~\onlinecite{kaib2020}. For simplicity and to reduce computational cost, we neglect their weak second-neighbor Dzyaloshinskii-Moriya and $J_2$ interactions. This model has the peculiarity that at $T=0, B=0$, the ferromagnetic state is lower in energy than the AF zigzag one on the classical level. The AF zigzag ground state is nevertheless recovered in exact diagonalization through strong quantum fluctuations. At zero and low field strengths, the lowest-energy excitation is accordingly found at one (or multiple) of the zigzag ordering wave vectors $ k=M,M',Y$, with the selectivity depending on the orientation of the magnetic field~\cite{Winter2018s}. In the conventional case (and for the minimal model), the excitations at such $k$ would still remain being the lowest-energy ones throughout the partially-polarized phase $B>\Bc[2]$, which can be anticipated already on the level of linear spin-wave theory. In the case of the present model however, zigzag correlations ($k=M,M',Y$) are only dominant over ferromagnetic ones $(k=\Gamma)$ with the help of strong quantum fluctuations. Since the latter are suppressed on approaching the high-field limit, the excitations related to ferromagnetic correlations eventually become energetically favoured. \Cref{fig:ShoulderModels}(a) shows the evolution of the 16 lowest-energy excitations calculated for this model. As anticipated, a crossing between the excitations of the $k=Y$ sector and the $k=\Gamma$ sector takes place in the partially polarized phase. The associated abrupt change of the slope of the gap ($\Delta'(B)$) leads to a jump in the zero-temperature limit of \GammaH\ and a shoulder-like anomaly at small finite temperatures [\cref{fig:ShoulderModels}(b)], that we associate with $\Bc[3]$ from experiment. At $T\gtrsim 1.5\,$K the shoulder becomes increasingly indistinguishable in this model. We stress that the overall agreement (aside from the existence of the shoulder) with experiment in this model without further adjustment is considerably poorer than in the minimal model for the temperature-evolution and absolute magnitude of most observables. In addition, while $\Bc[2]\approx 6\,$T is satisfied, the shoulder is rather far away from this critical field. 
 Nevertheless, we show with this as a proof of concept that in the parameter space of generalized Kitaev models, such excitation level crossings are common when the model is proximate to a phase that has lower energy than zigzag at high field strengths, but looses against zigzag at low field strengths. Then the low-field Hamiltonian has a zigzag ground state and the lowest-energy excitations remain to be of that nature in the partially-polarized until $\Bc[3]$, yet they eventually get undercut by the excitations related to the proximate phase above \Bc[3].     
In the case of this model, the proximate phase is likely the ferromagnetic one. Further, we have verified that the two lowest-energy excitations [purple- and yellow-colored in Fig. S8a] carry large intensity in the dynamical spin-structure factor. Therefore, measurements of the inelastic neutron scattering (INS) dispersion should observe that the location of the softest mode in k space switches around \Bc[3]. We speculate that the best way to observe this in INS would be to compare the dispersion at $B$ slightly above \Bc[2] with the dispersion at fields $B\gg \Bc[3]$. In practice, this might be a rather subtle effect to observe in INS since for fields $B > \Bc[2]$ the dispersion's bandwidth might be smaller than the overall gap. Alternatively it might be worthwile to measure the dispersion \textit{at} \Bc[3] (instead of comparing $B<\Bc[3]$ and $B>\Bc[3]$) whether there is a wave vector $k$ that is approximately degenerate with the ordering wave vector of the AF2 phase. Note that the the high-field lowest-energy $k$ point does not necessarily have to be $k=\Gamma$ as in our toy-model example. Instead it could also be a different wave vector in the three-dimensional Brillouin Zone.

\begin{figure}
	\includegraphics[width=0.9\linewidth]{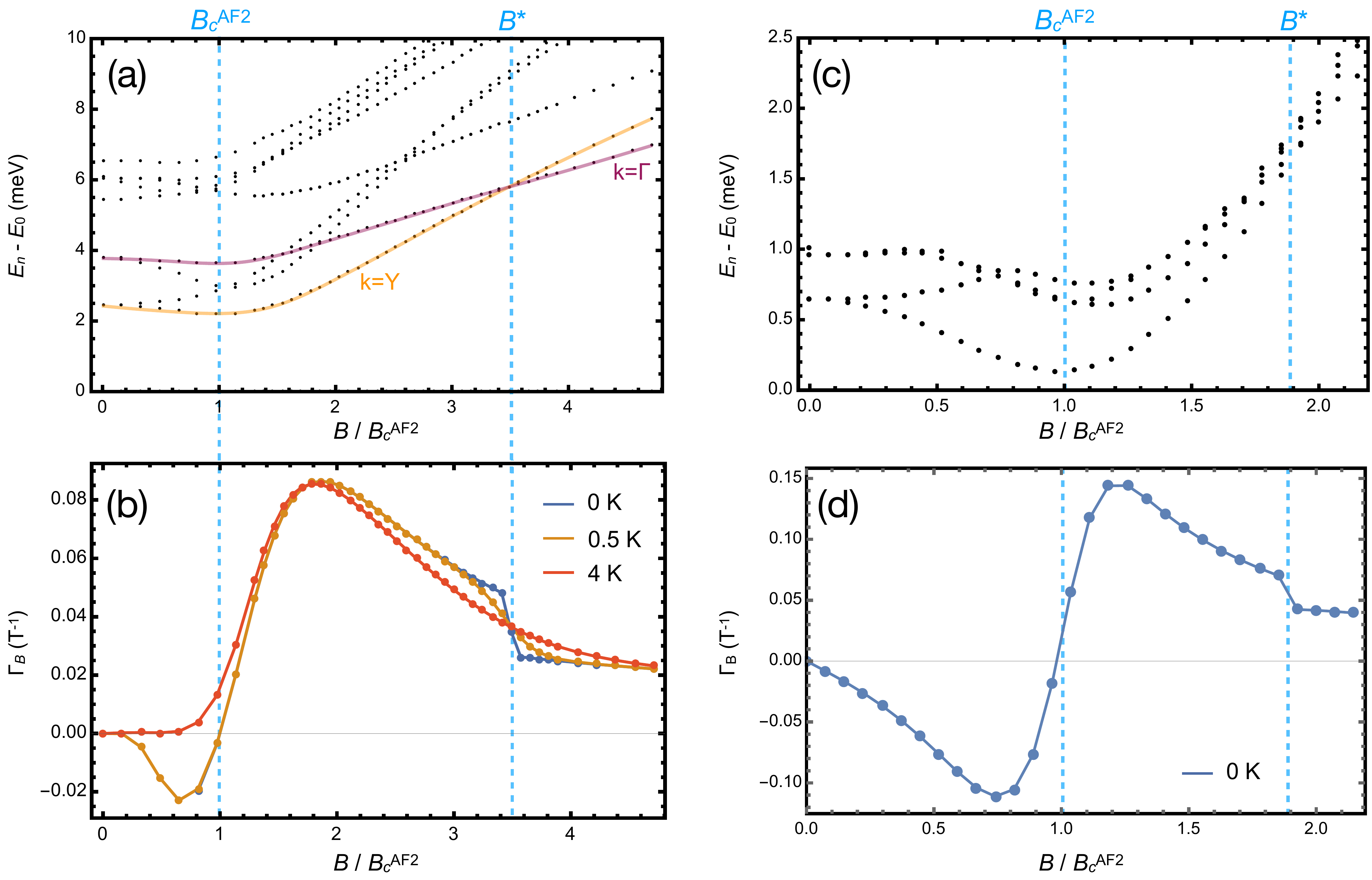}
	\caption{(a,b) Field-dependence of the lowest 16 eigenenergies and \Gruneisen\ parameter of the \textit{ab-initio} model of Ref.~\onlinecite{kaib2020} (neglecting weak second-neighbor Dzyaloshinskii-Moriya and $J_2$ couplings). We find $\Bc[2]\approx 6\,$T for this model. 
	(c,d)~Field-dependence of the lowest 6 eigenenergies and $T=0$ \Gruneisen\ parameter for a model interpolated between the minimal model~\cite{Winter2017bs} and the hidden AF Kitaev point~\cite{kaib2019} with parameters $(J_1,K_1,\Gamma_1,\Gamma'_1,J_3)=(3.35,-3.05,3.65,-1.7,0.05)\,$meV, $g_{ab}=2.3$. Here $\Bc[2]\approx 26\,$T.
	\label{fig:ShoulderModels}}
\end{figure}

 In order to test our hypothesis that vicinity to other phases may induce such excitation level crossings and in consequence shoulder-like anomalies in \GammaH, we tuned the original minimal model~\cite{Winter2017bs} closer to various phase boundaries and found further examples. By introducing large $J_1>0$ and $K_2<0$, we even find models with \textit{two} shoulder-anomalies above $\Bc[2]$. We note that proximity to a phase alone is not sufficient, as the adjacent phase may not become more attractive or even become less attractive by the magnetic field. Such a case is found when introducing large $\Gamma'_1>0$, where the adjacent phase is likely the incommensurate one. An exotic scenario one could imagine is that the adjacent phase is a quantum spin liquid (QSL). While we could not find shoulder-anomalies by tuning towards the ferromagnetic Kitaev point ($K=-1$), we found them by tuning the minimal model towards the \textit{hidden} AF Kitaev point~\cite{kaib2019}, whose nonzero parameters are $(J_1,K_1,\Gamma_1,\Gamma'_1)=(\frac49,-\frac13,\frac49,-\frac29)$. This model is dual to the AF Kitaev model $(K_1=1)$ and therefore hosts both the Kitaev QSL at low fields and a field-induced $U(1)$-QSL at intermediate fields~\cite{hickey2019emergence}. The hidden dual point is interesting since it is nearer in parameter space to realistic \aRuCl\ models as established by \textit{ab-initio} due to $K_1<0$, $\Gamma_1>0$, $\Gamma'_1<0$. This allows a simple path in parameter space where the \aRuCl\ models' zigzag ground states border the Kitaev QSL ground state: $H(g)=g H_\text{hidden Kitaev} \cdot 8.5\text{meV} + (1-g) H_\text{minimal model}$, where $g\in[0,1]$ and the subscripts refer to the hidden AF Kitaev point and the minimal model of Ref.~\onlinecite{Winter2017bs} respectively. At zero field, the QSL then borders the zigzag ground state at $g\simeq 0.04$. We find a level crossing of the lowest excitations in the field-induced partially-polarized phases above zigzag for $0.04<g\lesssim 0.1$. An example is shown in \cref{fig:ShoulderModels}(c,d), where $g=0.1$. Note that we can here only show with confidence the $T\rightarrow 0$ limit since we only calculated the 6 lowest-energy states as we are only interested in the two lowest-energy excitations above the ground state. These calculations indicate that the adjacent phase in parameter space inducing the shoulder anomaly \Bc[3] could also be a QSL. We note that this model is again to be understood as a toy model for a proof of concept, since aside from the existence of a shoulder-anomaly, many other properties of \aRuCl\ are not reproduced without further adjustment.

\section{Zero-temperature limit of the Gr\"uneisen parameter of gapped spectra}
For any continuous spectrum with a gapped ground state, the density of states can be written in the form 
\begin{equation}
  \text{DOS}(E,\control) = \delta(E) + \Theta\left[E-\Delta(\control)\right] f(E,\control), \label{eq:CutoffFormula}
\end{equation}
where $\Theta(x)$ is the Heaviside function, $f(E,\control)$ is any function describing the density of states above $\Delta>0$ and $\control$ is a control parameter like magnetic field ($\control=B$) measured in this study or external pressure ($\control=p$). The partition function is then given by
\begin{equation}
  Z = \int_{-\infty}^\infty   \text{DOS}(E,\control) e^{-\frac ET}\, \mathrm dE = 1+ \int_{\Delta(\control)}^\infty   f(E,\control) e^{-\frac ET}\, \mathrm dE,
\end{equation}
where we set $k_B=1$ throughout the derivation. With $F=- T \log Z$, $S=-\partial F / \partial T$, the Gr\"uneisen parameter associated to $\control$ can be straightforwardly written out as
\begin{align}
\Gamma_\control =&  \frac{(\partial S / \partial \control)}{T(\partial S / \partial T)}\\
=& \frac{T \left[\int_{\Delta}^{\infty } e^{-\frac{E}{T}} \partial_\control f (E,\control) \,\mathrm d E \left(TZ-\int_{\Delta}^{\infty } E e^{-\frac{E}{T}} f(E,\control) \,\mathrm d E\right)+Z \int_{\Delta}^{\infty } E e^{-\frac{E}{T}} \partial_\control f(E,\control) \,\mathrm d E\right]}{\left(\int_{\Delta}^{\infty } E e^{-\frac{E}{T}} f(E,\control) \,\mathrm d E\right)^2-Z \int_{\Delta}^{\infty } E^2 e^{-\frac{E}{T}} f(E,\control) \,\mathrm d E} \\
&+\frac{T\left[\Delta '(\control) e^{-\frac{\Delta}{T}} f(\Delta,\control) \left(\int_{\Delta}^{\infty } E e^{-\frac{E}{T}} f(E,\control) \,\mathrm d E-Z (\Delta+T)\right)\right]}{\left(\int_{\Delta}^{\infty } E e^{-\frac{E}{T}} f(E,\control) \,\mathrm d E\right)^2-Z \int_{\Delta}^{\infty } E^2 e^{-\frac{E}{T}} f(E,\control) \,\mathrm d E},
\end{align}
where  $\Delta'(\control)\equiv \partial \Delta/ \partial \control $ and $\partial_\control f(E,\control) \equiv \partial f(E,\control) / \partial \control$. 
For the $T\rightarrow0^+$ limit, we can neglect terms in nominator and denominator that are quadratic in $e^{-E/T}$ for all $E \geq \Delta > 0$ compared to terms that are linear in $e^{-E/T}$ as long as the latter do not cancel to zero (shown below). This leads to
\begin{align}
	\lim_{T\rightarrow 0^+}  \Gamma_\control 
	&=
		\lim_{T\rightarrow 0^+} \frac{  f(\Delta,\control) \Delta'(\control)(T+\Delta)
	 -Te^{\Delta/T} \int_\Delta^\infty \partial_\control f(E,\control) e^{-\frac{E}{T}} \mathrm d E  -
	 e^{\Delta/T}   \int_\Delta^\infty \partial_\control f(E,\control) E e^{-\frac{E}{T}} \mathrm d E}
	{T^{-1} e^{\Delta/T} \int_\Delta^\infty  f(E,\control)E^2 e^{-\frac{E}{T}} \mathrm d E} \label{eq:derivationfrac1}
	\end{align}
In order to form the low-temperature limit, it is useful to dissect the terms of the form $e^{\Delta/T} \int_\Delta^\infty a(E,\control) \mathrm dE$ into powers of $T$, which is possible by repeated partial integration:
	\begin{align}
		 e^{\frac\Delta T} \int_\Delta^\infty a(E,\control) e^{-\frac{E}{T}} \mathrm d E
		&= 
		Ta(\Delta,\control) + Te^{\frac\Delta T}   \int_\Delta^\infty  \frac{\partial a(E,\control) }{\partial E} e^{-\frac{E}{T}}\mathrm d E \\
		&=	Ta(\Delta,\control) + T^2 \left. \frac{\partial^1 a(E,\control) }{\partial E^1} \right|_{E=\Delta} + T^3 \left. \frac{\partial^2 a(E,\control) }{\partial E^2} \right|_{E=\Delta} + (\dots)
		,
	\end{align}
	where we used that for all three cases of $a(E,\control)$, it grows sub-exponentially with $E$ in the limit of large $E$. Inserting this for the three cases of $a(E,\control)$ and keeping track of the powers in $T$, one arrives at 
	\begin{equation}
  \lim_{T\rightarrow0^+} \Gamma_\control =\frac{ \Delta'(\control)}{\Delta(\control)}. \label{eq:dgapgap}
\end{equation}

It has been argued that if a system is dominated by one energy scale $E^*$, the \Gruneisen\ ratio follows $\Gamma_\control = (\partial E^\ast / \partial \control)/E^\ast$. For a gapped system and temperatures much below the gap one may anticipate that this energy scale becomes the gap. In this sense, our derivation above is a proof of this identification $E^\ast=\Delta$ in the $T\rightarrow 0$ limit.

\Cref{eq:dgapgap} implies that if the gap closes and reopens as $\Delta(\lambda)\propto |\lambda-\lambda_c|^p$ with any power $p$ at a critical $\lambda_c$, the \Gruneisen\ parameter diverges and changes sign as $\Gamma_{\lambda}(T\rightarrow 0) \propto (\lambda-\lambda_c)^{-1}$. The discontinuous behavior for a level crossing in the lowest-energy excitations was illustrated in the main text. We give below a short proof that $\Gamma_{\lambda}(T\rightarrow 0)$ changes sign and diverges for \textit{any} form of the gap closing and reopening.

 $\Delta(\lambda)$ having a minimum at $\lambda=\lambda_c$ implies that there exists an $s>0$ so that for every $r\in(0,s)$, $\Delta(\lambda)$ is monotonically growing for $\lambda_c<\lambda<\lambda_c+r$. In this interval, the only candidate for a pole of $\frac{\Delta'(\lambda)}{\Delta(\lambda)}$ is therefore at $\lambda\rightarrow \lambda_c^+$. We prove now that such a pole exists in this interval by showing that the integral over $(\lambda_c,\lambda_c+r)$ diverges:
 \begin{align}
 	\int_{\lambda_c^+}^{\lambda_c+r} \frac{\Delta'(\lambda)}{\Delta(\lambda)} \mathrm d\lambda = 
\left[\log(\Delta(\lambda))\right]_{\lambda_c^+}^{\lambda_c+r} = \log(\Delta(\lambda_c+r)) - \lim_{\mu\rightarrow \lambda_c^+} \log(\Delta(\mu)) = \log(\Delta(\lambda_c + r)) - \lim_{\mu\rightarrow 0^+} \log(\mu) = +\infty. \label{eq.divergence1}
 \end{align}
 Analogously for approaching $\lambda_c$ from below, there exists an $m >0$ so that for every $r\in(0,m)$, the gap $\Delta(\lambda)$ falls monotonically in the region $(\lambda_c -r, \lambda_c)$, and
  \begin{align}
 	\int_{\lambda_c-r}^{\lambda_c^-} \frac{\Delta'(\lambda)}{\Delta(\lambda)} \mathrm d\lambda = 
\left[\log(\Delta(\lambda))\right]_{\lambda_c-r}^{\lambda_c^-} =\lim_{\mu\rightarrow \lambda_c^-} \log(\Delta(\mu)) -  \log(\Delta(\lambda_c -r)) = \lim_{\mu\rightarrow 0^+} \log(\mu) -  \log(\Delta(\lambda_c -r)) = -\infty. \label{eq.divergence2}
 \end{align}
 Since $r$ in \cref{eq.divergence1,eq.divergence2} can be taken arbitrarily close to zero, the pole must be at $\lambda\rightarrow\lambda_c^-$. Taken together, these results show the divergence and sign change behavior of $\Gamma_\lambda(T\rightarrow 0)$ at a gap closing and reopening of any form. 

\end{widetext}

%

\end{document}